\documentstyle[preprint, revtex, eqsecnum]{aps}

\newcommand{\be}{\begin{equation}}
\newcommand{\ee}{\end{equation}}
\newcommand{\bea}{\begin{eqnarray}}
\newcommand{\eea}{\end{eqnarray}}
\newcommand{\iii}{\mbox{$ | i \rangle$}}
\newcommand{\fff}{\mbox{$ | f \rangle$}}

\newcommand{\sqiii}{\mbox {$\sqrt{3} \times \sqrt{3}$}}
\newcommand{\hh}{\mbox{$ {\cal H} \! \!$ }}
\newcommand{\whh}{\mbox{$ \widehat{{\cal H}} \! \!$ }}
\newcommand{\DD}{\mbox{$ {\cal D} \! \! $ }}
\newcommand{\LL}{\mbox{$ {\cal L} \!\! $ }}

\newcommand{\wLL}{\mbox{$ \widehat{{\cal L}} \!\!$ }}
\newcommand{\NN}{\mbox{$ {\cal N} \!\!$ }}
\newcommand{\SS}{\mbox{$ {\cal S} \!\!$ }}
\newcommand{\SSo}{\mbox{$ {{\cal S} }_o \!\! $ }}
\newcommand{\bsp}{\mbox{\boldmath $s$}}

\newcommand{\bon}{\mbox{\boldmath $n $}}
\newcommand{\bor}{\mbox{\boldmath $r $}}
\newcommand{\boz}{\mbox{\boldmath $z $}}
\newcommand{\whlphi}{\mbox{$\overline{\Phi}$}}
\newcommand{\whlz}{\mbox{$\overline{Z}$}}
\newcommand{\whz}{\mbox{$Z$}}
\newcommand{\whphi}{\mbox{$\Phi$}}

\newcommand{\lz}{\mbox{$\overline{z}$}}

\newcommand{\dz}{\mbox{$\delta z$}}
\newcommand{\whdz}{\mbox{$\delta Z$}}
\newcommand{\lphi}{\mbox{$\overline{\varphi}$}}
\newcommand{\dphi}{\mbox{$\delta \varphi$}}
\newcommand{\whdphi}{\mbox{$\delta \Phi$}}
\newcommand{\llz}{\mbox{$\overline{Z}_c$}}
\newcommand{\Zc}{\mbox{$Z_c$}}
\newcommand{\delz}{\mbox{$\delta \! z$}}

\newcommand{\toh}{\mbox{$\textstyle{ \frac{1}{2}}$}}

\newcommand{\unz}{\underline{0}}
\newcommand{\unpi}{\underline{\pi}}
\newcommand{\past}{\varphi^{\ast}}
\newcommand{\lrho}{\overline{\rho}}
\newcommand{\supx}{\rm \scriptscriptstyle{x}}
\newcommand{\supy}{\rm \scriptscriptstyle{y}}
\newcommand{\supz}{\rm \scriptscriptstyle{z}}
\newcommand{\boe}{\mbox{\boldmath $e $}}
\newcommand{\widef}{\widetilde{f}}
\newcommand{\wJ}{\widetilde{J}}
\newcommand{\lphiT}{\overline{\varphi}_{\tau} }

\tightenlines

\begin{document}
\columnsep0.1truecm
\draft
\preprint{ }
\begin{title}
Spin tunneling in the Kagom\'e antiferromagnet
\end{title}
\author{Jan von Delft and Christopher L. Henley}
\begin{instit}
Laboratory of Atomic and Solid State Physics,
Cornell University, Ithaca, NY, 14853
\end{instit}
\receipt{ }
\begin{abstract}
The collective tunneling of a small cluster of spins
between two degenerate ground state configurations of the
Kagom\'{e}-lattice quantum Heisenberg antiferromagnet is \mbox{studied}.
The cluster consists of the six spins on a hexagon of the lattice. The
resulting tunnel splitting energy $\Delta$ is calculated in detail,
including the prefactor to the exponential $\exp(- \SSo / \hbar)$. This is
done by setting up a coherent spin state path integral in imaginary time
 and evaluating it by the method of steepest descent.
The hexagon tunneling problem is mapped onto a much simpler
tunneling problem, involving only one collective degree of freedom, which
can be treated by known methods. It is found that for half-odd-integer
spins, the tunneling amplitude and the tunnel splitting energy are exactly
zero, because of
destructive interference between symmetry-related
 $(+)$-instanton and $(-)$-instanton tunneling paths. This destructive
interference is shown to occur also for certain larger loops of spins on
the Kagom\'{e} lattice. For small, integer spins, our results suggest
that tunneling strongly competes with
\mbox{in-plane} order-from-disorder selection effects;
it constitutes a disordering mechanism that might drive the system
into a partially disordered ground state, related to a spin nematic.
\end{abstract}


\narrowtext

\section{Introduction}

\label{intro}
\indent
Consider a Heisenberg antiferromagnet (AFM) with non-random but
competing exchange interactions. The classical
ground state is often non-trivially
non-unique, in having a continuous manifold of inequivalent (but degenerate)
ground states. However, if one takes account of quantum and thermal
fluctuations around the classical ground states, the nontrivial degeneracy
may be broken. The effects
of fluctuations can generally be represented by an effective ``selection''
Hamiltonian,  which is a function of the classical spin directions
and ``selects'' certain ground states (sometimes having long-range order)
in favor of  others. Since long-range
order can thus be induced out of an apparently disordered manifold of
ground states,
such selection effects are called ``ordering due to disorder'' \cite{Vil80}.

An effect that {\em competes}\/ with ``order from disorder'' selection
effects is tunneling between different ground state configurations.
Tunneling tends to drive the system into a superposition of degenerate
states, rather than selecting a particular one. Hence,
in the regions of parameter space in which tunneling events are important,
they could suppress ``order from disorder'' selection effects.

In this paper we study spin tunneling
in a 2D quantum Heisenberg antiferromagnet
on a Kagom\'e lattice \cite{Els89,HKB92,RCC92,Sac92}. This is a
frustrated spin system with a very large ground state degeneracy, in which
various  selection effects have been investigated
\cite{RCC92,Sac92,CHS92,CH92,Chu92}.
We study tunneling events that involve the rigid simultaneous
rotation of small groups of spins (``weathervane modes'',
defined in section~\ref{selection}, in particular a mode involving only the
six spins on a Kagom\'e hexagon).
 We find, rather unexpectedly,  that the tunneling amplitudes
are zero when the spin
$s$ is half-odd-integer, but non-zero when $s$ is integer.
This is due
to destructive interference between
two topologically distinct tunneling paths connecting the same initial and
final states. The interference occurs when the tunneling
amplitudes have different topological phase factors \cite{vDH92a}.
Therefore there should be interesting integer vs. half-odd-integer $s$
effects for that range of $s$-values for which tunneling effects are
as strong or stronger than selection effects: systems with
integer $s$ would have a greater tendency to
be disordered because tunneling suppresses
selection effects, whereas systems with half-odd-integer $s$,
where tunneling is absent, would tend to be ordered.

Apart from studying the role of tunneling in the Kagom\'{e} lattice, we
hope that this paper will provide an instructive example of a rather
non-trivial spin-tunneling calculation. As is customary, the tunneling
amplitude of interest is calculated by setting up a coherent-spin-state path
integral in imaginary time
 and evaluating it by the method of steepest descent \cite{Kla79}, which
is an expansion in powers of $1/s$. Our calculation includes a
complete evaluation of the prefactor to the exponential $e^{- \SS_o/ \hbar}$
($\SS_o$ is the classical action),
and a discussion of the integration ranges of the spin path integral (these
ranges are finite originally, but need
to be extended to infinity to allow an evaluation of the prefactor).
Although the the  calculation of  tunneling rates for spin systems has
 been of interest in various different contexts
 \cite{CG88,GK90,AL91,ASG92,KG92}, we are aware of only one recent
paper where such prefactors are calculated explicitly
\cite{KG92}. Moreover, we show explicitly how one may give an exact
treatment of the simultaneous collective motion of all the relevant spin
 degrees of freedom, by reducing the problem to one involving only a
single,  collective degree of freedom. The reduced problem
can be treated by methods
well-known from studying a particle in a  double-well potential.

The results of our calculations suggest that for small, integer $s$, the
tunneling amplitude is sufficiently large that tunneling can be regarded as
a significant disordering mechanism, that tends to drive the system into a
partially disordered state, related to a spin-nematic state
\cite{CHS92,CC91}.

This paper is organized as follows: In section~\ref{selection} we review
ground state selection effects on the Kagom\'e lattice, and describe the
hexagon tunneling event that is to be studied in later sections.
In sections~\ref{mod} and~\ref{tunnelsplit}
we study a simple model Lagrangian, chosen such that it will be of use for the
subsequent study of the Kagom\'{e} system. Specifically, the calculations
of the classical
action $\SS_o$ (section~\ref{mod}) and the tunnel splitting energy $\Delta$
(section~\ref{tunnelsplit}) are presented in detail.
 Section~\ref{Kag} shows how the full hexagon
tunneling problem can be mapped onto the simple model problem studied
in earlier sections, and contains the main result of this paper,
eq.~(\ref{eq86.9}). In section~\ref{desint} the occurrence of
destructive interference during the tunneling of larger sets of spins on the
Kagom\'{e} lattice is discussed. There are four appendices.
In appendix~\ref{appD} an estimate is made of the
size and shape of the coplanarity potential that we employ in
section~\ref{Kag} to study the hexagon tunneling problem.
Appendix~\ref{appB} contains a summary of results that are
well-known for a particle tunneling in a double well and are needed in
section~\ref{tunnelsplit}.

\section{The Kagom\'e antiferromagnet}

\label{selection}
The Kagom\'{e} lattice is a 2D triangular
lattice with lattice constant $a_o$, where sites have been removed
at all sites of a triangular superlattice, with lattice constant $2a_o$ (see
fig.~\ref{ABABAB}).
The quantum Heisenberg AFM on this lattice has the Hamilton operator
\begin{equation}
\label{kagham}
\widehat{\cal H} =  J \sum_{ \langle i,j \rangle}
\widehat{\bsp}_i \! \cdot \!  \widehat{\bsp}_j \; ,  \qquad (J > 0) \;
\end{equation}
where $\widehat{\bsp}_i$ is the spin vector operator for the $i$-th spin and
the sum runs over all nearest neighbors.
Coherent spin states (see, e.g.\ \cite{Lieb})
may be used to discuss this system in classical terms.
Associate the coherent spin state $| \varphi_i ,\theta_i \rangle$
with the $i$-th spin, where the spherical coordinates
$(\varphi_i,\theta_i) \equiv \Omega_i$
define a unit vector $\bon_i$ (fig.~\ref{n,triangle}a).
The well-known property $ \langle \phi_i, \theta_i |
\widehat{{\mbox{\boldmath $s $}}}_i | \phi_i,
\theta_i \rangle = s {\mbox{\boldmath $n $}}_i$, allows one to interpret
$\bon_i$ as a ``classical spin vector''.
Also, a ``classical''
Hamiltonian $\hh$ may be introduced as the coherent-spin-state
expectation value $\langle \whh \rangle$:
\begin{equation}
\label{hhexp}
{\cal H} \equiv \left( \prod_{i \otimes} \langle \phi_i, \theta_i | \right)
\whh \left( \prod_{i \otimes} | \phi_i, \theta_i \rangle \right) \, = \,
 s^2 J \sum_{ \langle i,j \rangle}
\bon_i \! \cdot \! \bon_j \; ,  \qquad (J > 0) \;
\end{equation}
 When referring to the ``energy'' of a state, we shall always mean this
expectation value $\hh$. Likewise, the term ``ground state'' will not be
used to refer to an actual eigenstate of the operator
$\whh$, but to a state that
minimizes $\hh$. It is this $\hh$ that is used in discussions of the
classical Kagom\'e antiferromagnet.

The energy is minimized by any configuration
 in which the total spin on {\em each}\/ elementary
triangle of the lattice is zero \cite{Els89,CHS92}.
In such states, the spins of any given triangle of the Kagom\`e lattice
lie in one plane, with
relative angles of $120^{\circ}$ (see fig.~\ref{n,triangle}b), forming a rigid
unit in spin space. Because of the many ways of fitting such triangles
together,  the distinct classical ground states
form a manifold with a dimensionality proportional to the system size.

\subsection{Selection effects}

\label{selectioneffects}
It is convenient to describe the classical ground states with
reference to a coplanar ground state, in which all spins lie in the same
 ``reference plane''.
All the coplanar ground states are constructed by labelling
the sites by letters A,B and C, such that each triangle has one of each, and
then replacing the three kinds of letters by spins in three directions
differing by $120^\circ$ angles \cite{Els89} (fig.~\ref{n,triangle}b);
these states correspond one-to-one to the ground states
 of a 3-state Potts AFM on
a Kagom\'e lattice.

All non-coplanar classical
states can be generated by continuous distortions of
coplanar   states, henceforth to be called ``Potts states'',
without crossing energy barriers \cite{CHS92}.
For example, any hexagon of six spins that is labelled only by two letters
(e.g.\ ABABAB in fig.~\ref{ABABAB})
can be rotated as a rigid unit in spin space by an angle $\varphi$ around
the C-direction without
any cost in classical energy; this has been called  the ``weathervane'' mode
by Ritchey {\em et.\ al.}\/ \cite{RCC92}.
However, by expanding in spin waves about any given ground state, it has
been shown that all non-coplanar states have a larger zero-point
energy than the coplanar
Potts states. This selection effect can be characterized by
a parameter $J_b \sim O(sJ)$, and ensures that
the true ground state will be  coplanar \cite{RCC92,Sac92,CHS92,CH92,Chu92}.

Further study, using a self-consistent approach, yields a smaller
``in-plane'' selection energy, of order $J_c \sim O (s^{\frac{2}{3}} J)$,
  which favors among all possible
Potts states a particular state with long-range order, the so-called
\sqiii\ ground
state \cite{Sac92,CH92,Chu92}, depicted in fig.~\ref{ABABAB}.
 In this state, {\em every}\/ hexagon is labelled by just
two letters, e.g.\ BCBCBC. To encode the most important effects of this
``in-plane'' selection, it is convenient to define a ``chirality'' variable
centered on each triangle, equal to $-1$ or $+1$ depending on whether the sites
take values ABC in the clockwise or counterclockwise sense.
In the \sqiii\ Potts state, the chiralities on neighboring
elementary triangles are antiferromagnetically ordered. Hence the selection
energy can be expressed as
 an effective antiferromagnetic  coupling, of strength $J_c$, say,
 between the chiralities.

\subsection{Competition between selection effects and tunneling}

The above considerations
 neglected the possibilities of large-scale fluctuations and
tunneling between classical ground states.
Clearly the smallest object which can tunnel is the ``weathervane mode'':
Consider as initial state, \iii, the \sqiii\ Potts state.
If the six spins on an ABABAB-hexagon rotate by $180^{\circ}$ around the
C-direction (take the latter to define the $\boz$-axis), another Potts state,
\fff, with a BABABA-hexagon is reached (fig.~\ref{BABABA}).
 In the absence of in-plane selection ($J_c=0$), \iii\ and \fff\
would be degenerate.  Tunneling
between  \iii and \fff\ would tend to drive the system
into a superposition $\frac{1}{\sqrt{2}} ( \iii \pm \fff)$, with energy $E_o
\pm \Delta$ (where $\Delta$ is proportional to the tunneling amplitude).
Now, in the presence of in-plane selection effects,
the energies of \iii\ and \fff\ differ by $12J_c$, since the chiralities on
all six triangles bordering the hexagon of \iii\ are opposite to those of
\fff.  If $12 J_c > \Delta$,
 the tunneling amplitude is very small, and
the system selects \iii, the \sqiii\ state.
If, on the other hand, $12 J_c < \Delta$ and the tunneling amplitude is
appreciable, the system
can lower its energy by adopting the superposition $\frac{1}{\sqrt{2}} (
\iii - \fff)$.

Such a  hexagon tunneling event can clearly take place starting from any Potts
state
\iii\ that contains an ABABAB-type hexagon, not just from the \sqiii\ state.
Thus, tunneling constitutes a disordering factor that competes with in-plane
selection. If such hexagon tunneling events occur with large probability
throughout the Kagom\'e lattice, a  ground state
will result that is at most partially ordered (related to a spin nematic)
(see section \ref{discussion}).

\subsection{Effective hexagon Hamiltonian}

\label{hte}

We shall study the hexagon tunneling event described in the previous section
 in the absence
of in-plane selection (i.e.\ taking $J_c=0$), but in the presence of
a coplanarity barrier of size $J_b$, that tends to keep the spins aligned
with the reference plane. Our aim is to compute the tunnel
splitting energy $\Delta$. Since $J_b \sim s J$, we have $J_b \ll s^2 J$,
when $s \gg1$, and consequently we assume
that only  the six hexagon spins
 will move significantly during the hexagon tunneling event.
Hence we take all other spins in the lattice to stay fixed, and adopt the
following Hamiltonian for the six-spin hexagon system:
\be
\label{bigh}
\hh_{\rm hex}= \hh_{\rm AFM} + \hh_{\rm cop} \; ,
\ee
\bea
\label{AFM}
{\rm where} \qquad
\hh_{\rm AFM} & = &  s^2 J \sum_{l = 0}^5 \bigl[
 \bon_l \! \cdot \!
\boz + \bon_{l+1} \! \cdot \!  \boz +
\bon_l \! \cdot \! \bon_{l+1} + {\textstyle \frac{3}{2}} \bigr] \; , \\
\label{cop}
\hh_{\rm cop} & = & J_b \, f( \varphi_{\rm av}) \; , \quad {\rm where}
\; \; \varphi_{\rm av} \equiv 1 / 6
\sum_{l = 0}^5 (\varphi_l - \varphi_l ({\rm {\scriptstyle initial}}) ) \; .
\eea
The index $l$ labels the six spins around the hexagon and is defined modulo
6 (see fig.~\ref{BABABA}).
All spins are written in terms of the same coordinate system, in which the
$\boz$-axis points in the C-direction, the $xz$-plane coincides with the
reference plane, and $\varphi$ is measured from the positive $x$-direction.
$\hh_{\rm AFM}$ contains simply those terms from eq.~(\ref{hhexp})
that involve the six hexagon spins and their six C-type nearest neighbors.
A constant has been added to ensure that $\hh_{\rm AFM}
 = 0$ in the initial and final configurations $|i \rangle$ and $| f \rangle$.

$\hh_{\rm cop}$ represents the coplanarity barrier that opposes hexagon
tunneling. The function  $f (\varphi)$
describes the shape of the barrier; it is of order unity and is
sketched schematically in fig.~\ref{barrier}. Its argument in eq.~(\ref{cop}),
 $\varphi_{\rm av}$, is a measure of
the deviation of the plane of the near-rigid hexagon
from the reference plane (in which $\varphi_{\rm B} (-T/2) = 0$,
$\varphi_{\rm A} (-T/2)= \pi$).

Symmetry about the reference plane ensures that
$f (\varphi)$ has the properties
\be
\label{copbarrier}
f ( m \pi ) = 0 \; , \qquad
 f (\varphi) = f (- \varphi) = f (\varphi + m \pi) \quad {\mbox {\rm
for any integer $m$}} \; .
\ee
At this stage it is not necessary to specify $f (\varphi)$ in more detail.
An estimate of the actual form of $\hh_{\rm cop}$ is made in
appendix~\ref{appD}, and summarized in eq.~(\ref{hcopf}).
The suitability of using such a barrier is further discussed in
section~\ref{discussion}.

Intuitively, the six hexagon spins are expected to rotate collectively,
almost as a rigid unit,
maintaining the  mutual coplanarity and relative angles of $120^{\circ}$
required to minimize $\hh_{\rm AFM}$. In particular, for each of them
$\cos \theta_l \simeq -\frac{1}{2}$ throughout the tunneling event,
which is why we have taken $\hh_{\rm cop}$
to be independent of $\theta_l$. Also, the expected mutual coplanarity of the
tunneling spins is the reason why we have written
$\hh_{\rm cop}$ as a function only of $\varphi_{\rm av}$ (and not some more
general function of the six $\varphi_l$'s \cite{moregeneral}).
These assumptions are found to be justified in section~\ref{So}.

Evidently, due to reflection symmetry in the
reference plane, two kinds of tunneling events between \iii\ and \fff\ are
possible. They  differ from each other
only in the {\em sense}\/ of rotation about the $\boz$-axis ($\varphi
\leftrightarrow - \varphi$), and we shall call them
$(+)$-instantons and $(-)$-instantons.

We shall show in section~\ref{Kag} how the hexagon
tunneling problem defined above can be mapped onto a much simpler
model problem. This involves only a single, collective spin degree of freedom
with effective spin $6s$. Its coordinates, to be denoted by
$(\Phi_o, \Theta_o)$, are formally defined in
eqs.~(\ref{eq79.2}) and~(\ref{eq79.3}) and are suitable averages of the
six individual $(\varphi_i, \theta_i)$'s. The effective Hamiltonian turns
out to be (see eq.~(\ref{eq83.7})):
\be
\label{modelham}
\hh_{\rm eff}  =  12 s^2 J (\cos \Theta_o + {\textstyle \frac{1}{2}})^2
 + J_b f( \Phi_o) \, .
\ee

Rather than proceeding with the mapping of eq.~(\ref{bigh}) onto
eq.~(\ref{modelham}) right away, in the next two sections
we first discuss the simple model problem (based on eq.~(\ref{modelham}))
in detail, to establish notation and introduce the tools needed
to calculate the tunnel splitting energy $\Delta$.

\section{A simple model system}
\label{mod}

In this section we introduce a simple model problem, which will be used to
illustrate how a calculation of the tunnel splitting energy can be carried
through. It is also the system onto which the Kagom\'{e} tunneling problem
that is studied in section~\ref{Kag} can be
mapped. We set up the relevant path integral, calculate the classical action
and discuss the equations of motion and the typical form of the tunneling
path.

\subsection{Model Hamiltonian}

\label{modham}
The model system is defined by the following Euclidean single-spin Lagrangian:
\bea
\label{eq87.1}
 \LL &=& - i \hbar n s \dot{\varphi} ( z - 1)  + \hh \\
\label{eq87.1a}
{\rm where} \qquad  \hh &= &  a(z - z_g)^2 + b f (\varphi) \; ; \qquad ({\rm
with}\; b \ll a )\; , \\
{\rm and} \qquad z &\equiv & \cos \theta \; .
\eea
This Lagrangian has been written in terms of the imaginary time parameter
$ \tau = i t$ (hence ``Euclidean"),
 since this is convenient for the calculation of
tunneling amplitudes.
The spherical coordinates $(\varphi,\theta) \equiv \Omega$ define a unit
vector $\bon$ and label a coherent
spin state $| \varphi,\theta \rangle$ for a particle with spin $ns$.
The dot on $\dot{\varphi}$ means $\partial_{\tau}$
(see \cite{Frad91} for a discussion of the origin of this term).
 The integer $n$ is introduced in order to accommodate the possibility
of a collective degree of freedom with effective spin $ns$. For the
purpose of describing a single spin degree of freedom, take $n = 1$.

The ``classical'' Hamiltonian is the expectation value
$\hh \equiv  \langle \varphi,\theta | \whh | \varphi,\theta \rangle$ of
the quantum operator $\whh$.  Clearly $\hh$ has been chosen to have the same
form as $\hh_{\rm eff}$ of eq.~(\ref{modelham});
the constants $a$ and $b$ are of order $s^2 J$ and $J_b$
respectively (see eq.(\ref{eq83.8}), with $b \ll a$.
The dominant term in $\hh$ is an easy-plane anisotropy which
would make every angle on the ``latitude line'' $z = z_g$ on the unit sphere
be degenerate;
it mimics $\hh_{\rm AFM}$ in eq.~(\ref{AFM}) (which forced all six hexagon
spins to have $\cos \theta_l = - \frac{1}{2}$).
In the other term, $f (\varphi)$
is the same function as that in eqs.~(\ref{cop}) and~(\ref{copbarrier});
 it introduces a small anisotropy within the degenerate subspace and mimics
$\hh_{\rm cop}$.
There are two degenerate ground state configurations,
$|i \rangle = | 2m_1 \pi, z_g \rangle $ and $|f\rangle  = |
(2m_2 + 1) \pi, z_g \rangle$, with $m_1$ and $m_2$ arbitrary integers
(the different values of $m_1$ and $m_2$ describe the same physical
state, of course). Contours of constant $\hh$ are shown in
fig.~\ref{contours}.

\subsection{Tunneling amplitude}

\label{tunnamp}
The tunnel splitting energy $\Delta$ that arises due to
tunneling  of the spin direction between \iii\ and \fff\
 is proportional \cite{Col85} to the tunneling amplitude
\be
\label{eqL1}
U_{fi} \, \equiv \, \langle f|
e^{- \widehat{\hh} T / \hbar} | i \rangle = \int \DD \Omega
e^{- \! \SS / \hbar} \; ,
\ee
where $\SS = \int_{- T/2}^{T/2} \! d \tau  \LL $ is the Euclidean
action, $\DD \Omega$ is the path integral
measure (discussed in section~\ref{intz}),
and $T$ is a large time.  Such an amplitude
can be approximately evaluated by the method of steepest descent:
\be
\label{eqL3}
U_{fi} \, = \sum_l \NN^{(l)} e^{- \SSo^{(l)} / \hbar } \, \equiv \,
U_{fi}^{(l)} \; .
\ee
Here
 $\SSo^{(l)}$ is the action evaluated
along the $l$-th ``tunneling path'', which is a solution
to the Lagrangian equations of motion and for tunneling problems
 is in general complex. It will always be denoted by overlined variables,
e.g. $(\lphi^{(l)}, \lz^{(l)})$. The index
$l$ allows for the possibility of different tunneling paths satisfying
 boundary conditions that differ in the indices $m_1$ and $m_2$ (but that
all describe physically the same initial and final states) \cite{boundaries}.
The prefactors
\be
\label{eqL3a}
\NN^{(l)} = \int \! \! \DD \Omega  e^{- (\SS - \SS_o^{(l)})  / \hbar} \; .
\ee
are usually evaluated only (if at all) to lowest order in the
steepest descent method, by transforming to the fluctuation
variables $(\dphi^{(l)}, \dz^{(l)}) \equiv (\varphi - \lphi^{(l)}, z -
\lz^{(l)})$ and keeping only the lowest term in the expansion
$\SS - \SS_o^{(l)} = \delta^2 \SS^{(l)} + \delta^3 \SS^{(l)} \dots$.

In the present case, all tunneling paths connecting $|i \rangle$
 at $\tau = - T/2 $ to $| f \rangle$ at $\tau =  T/2$, can be
constructed from two very simple paths, to be denoted by $(\lphi^{\pm},
\lz^{\pm})$. The first, for which $\lphi^{+} (\tau) \in [0, \pi]$, we
 call a $(+)$-instanton, the second, for which $\lphi^{-} = - \lphi^{+} (\tau)
\in [ 0, - \pi ]$, a $(-)$-instanton. They differ solely in the sense
of $\varphi$-rotation and are sketched in fig.~\ref{instantons}. All
other tunneling paths that approximately satisfy the equations of motion
and contribute to eq.~(\ref{eqL3})
are multi-instanton paths. They consist of $n_1$ $(+)$-instanton and $n_2$
$(-)$-instanton events, with $n_1 + n_2 = $~odd, all assumed widely
separated (relative to their characteristic width) in time, and
following each other in arbitrary order (this is the so-called dilute-gas
approximation, see Coleman \cite{Col85}). In the following, we consider only
single-instanton events (i.e. $l \rightarrow \pm$ in eq.~(\ref{eqL3}));
the effects multi-instantons are taken into account
 in  appendix~\ref{appB}.

The symmetry $\hh (\varphi, z) = \hh (-\varphi, z)$ of the Hamiltonian
allows one to conclude immediately that $|U_{fi}^+| = |U_{fi}^-|$. This
is intuitively obvious, and can be proven to hold exactly
to all orders of the steepest descent
approximation (i.e.\ to all orders in $1/s$) (see section~\ref{relphase}).
Intuitively speaking,
the symmetry of \hh\ ensures that the local neighborhoods of the two
tunneling paths $(\lphi^+, \lz^+)$ and  $(\lphi^-, \lz^-)$
 are identical for the two paths, so that
the local shapes and sizes of the barriers (which determine ${\rm
Re} [\SS_o^{(l)}]$) and the local fluctuations around the tunneling paths
(which determine $|\NN^{(l)}|$), are identical.  However, the amplitudes
$U_{fi}^{\pm}$ {\em can}\/ differ by a phase, which may lead to destructive
interference between them.

\subsection{Classical action, continuing to complex coordinates}

\label{classact}
Let us find $\SS_o^{\pm}$, the classical action for single-instanton
events. Since energy is conserved
along any path that extremizes the action, $\hh (\lphi^{\pm} , \lz^{\pm})
 = 0$. Solving for $\lz^{\pm}$ as a function of $\lphi$, one  obtains
\be
\label{eq7}
i \whlz^{\pm} \, \equiv \, \lz^{\pm} - z_g =  \pm i \sqrt{b/a} \,
\sqrt{ f (\lphi^{\pm})}  \; ,
\ee
with $\whlz^{\pm}$ real.
The Euclidean action is easily evaluated along these paths by changing
variables from $\tau$ to $\lphi^{\pm}$ and using $f (\lphi^+) = f (\lphi^-)$:
\bea
\nonumber
\SS_{o}^{\pm} & = &  -i  \hbar n s \int_{-T/2}^{T/2} \!\!\!
 d \tau (\partial_{\tau}
 \lphi^{\pm}) \,  (z_g + i \whlz^{\pm} -1) \;\, + \; \, 0  \\
\label{eq8}
& = & \mp  i \hbar n s \pi (z_g -1 ) +   \hbar n s \sqrt{b/a} \int_0^{\pi}
\!\! d \varphi \sqrt{ f (\lphi^+ )} \; .
\eea

Note that the quantity $\sqrt{b/a}$ plays the role of an effective barrier
size for the tunneling process. This is intuitively plausible.
The barrier height in the $\varphi$ direction is measured by $b$.
 When $(z - z_g)$ is real, $a$
measures the steepness of the valley in the $z$-direction. When $(z - z_g)$
becomes purely imaginary ($(z - z_g)^2 \rightarrow - (z - z_g)^2$),
the $z$-valley turns
into a ridge (see fig.~\ref{contours}).
 The motion occurs along a constant energy contour around
this ridge (see fig.~\ref{contours}b),
 and $1/a$ measures the width of the ridge. Hence, $b/a$
measures the effective barrier size.

\subsection{Possible cancellation due to phase factors}

\label{possphasediff}
We show in section~\ref{relphase} that $\NN^+ = \NN^-$ (under some technical
assumptions, explained there). The sum of the two amplitudes
$U_{fi}^{\pm}$ is therefore simply
\bea
U_{fi}^+ + U_{fi}^- &=& U_{fi}^+ (1 + e^{ \SS_o^+ - \SS_o^- ) / \hbar} )
\nonumber \\
\label{relphase1}
& = & U_{fi}^+ (1 + e^{- i \pi l} ) \; ,
\eea
where the constant $l \equiv 2 n s (z_g -1 )$
is not necessarily an integer. The relative phase of $e^{- i \pi l}$
between the amplitudes of
two tunneling paths that connect the same initial and final
states has a well-known geometrical interpretation. It is related to the
area enclosed on the unit sphere between the two paths
$(\varphi^+ ( \tau), {\rm Re}[\theta^+ (\tau)] )$ and
$(\varphi^- ( \tau), {\rm Re}[\theta^- (\tau)] )$.
(See \cite{vDH92a} for more comments on this aspect).
Now, {\em if $l$ is an odd integer},\/ the $(+)$-instanton and $(-)$-instanton
paths interfere
destructively and their amplitudes add to zero: $U_{fi}^+ + U_{fi}^-
 = 0$ \cite{multinst}. Note that as long
as the barrier does not violate the symmetry between $(+)$-instantons and
$(-)$-instantons, this cancellation does not depend on the particular shape
of the barrier, since the shape function $f(\varphi)$
only affects the real part of $\SS_o$, which cancels out in $\SS_o^+ -
\SS_o^-$.

\subsection{Equations of motion and tunneling path}

\label{tunnpath}
It was not necessary to solve the equations of motion to obtain
the  explicit expression eq.~(\ref{eq8})
for the classical action. For future reference, the equations of motion are:
\be
\label{eq87.3a}
- i \hbar n s   i \dot{\whlz}  \, = \, b f' (\lphi) \; ,
\ee
\be
\label{eq87.3b}
-i \hbar n s \dot{\lphi} + 2 a \,  i \whlz \, = \, 0 \; .
\ee
The prime on $f$ means $\partial / \partial \varphi$.
Note how the $i$ of $i \whlz$ and that originating from $\tau = i t$
combine to result in consistent equations of motion (this is a direct
illustration of why it is useful to employ imaginary time in tunneling
calculations). Eliminating $\whlz$, we get
\be
\label{eq9}
\ddot{ \lphi} = \toh d^2  f'( \lphi)  \; , \qquad {\rm where} \quad d^2 =
\frac{4 a b}{  (\hbar n s)^2} \; .
\ee
The constant $d$ has the dimensions of inverse time, and $1/d$ characterizes
the width of an instanton. To illustrate the nature of the solutions
of eq.(\ref{eq9}), consider the simple case where $f (\varphi) = \sin^2
\varphi$. Then eq.(\ref{eq9}) is just the sine-Gordon equation, whose
solutions, in the limit  $T \rightarrow \infty$, are:
\be
\label{eq10}
\lphi^{+} (\tau) = 2 \arctan ( e^{d t}) \, , \qquad \lphi^{-} (\tau)
= - \lphi^+ (\tau)  = \lphi^+ ( - \tau ) - \pi \; .
\ee
These represent $(+)$-instanton and $(-)$-instanton events
and are sketched in
fig.~\ref{instantons}. The shape of an instanton changes quantitatively,
but not qualitatively, if a different barrier shape function is used.

\section{The tunnel splitting energy $\Delta$}

\label{tunnelsplit}
For integer $s$, for which the $(+)$-instanton and $(-)$-instanton amplitudes
have the same sign and
interfere constructively, the resulting tunnel splitting energy $\Delta$
can be calculated from the prefactor of eq.~(\ref{eqL3a}):
\be
\label{eqL3ab}
\NN^{\pm} = \int \! \! \DD \Omega e^{- (\SS
- \SS_o^{\pm} )  / \hbar} \; .
\ee
 This can be accomplished via three steps: \newline
(i)  Extending the integration ranges for $\varphi$ and $z$ in the path
     integral to infinity; \newline
(ii) Integrating out the $z$-degree of freedom; \newline
(ii) Using standard methods to evaluate the resulting $\varphi$-path integral.

\subsection{Extending the integration ranges}

\label{intz}
The path integral expression eq.~(\ref{eqL1}) for the tunneling
amplitude (in real time) can be arrived at  by the usual procedure of
discretizing time and inserting completeness relations in spin space at
each time slice (see, e.g.~\cite{Frad91,Kla85}). This procedure leads to a
formal expression for the measure (appropriate for any spin problem):
\be
\label{eq88.8}
\int \DD \Omega (t) = {\rm lim}_{N \rightarrow \infty}
\left( \frac{2s +1}{4 \pi} \right)^N \prod_{j=1}^{N}
\int_o^{2 \pi} \! d \varphi ( \tau_j) \int_{-1}^1 \! d z(\tau_j)  \; ,
\ee
where $\varepsilon (N + 1) = T$. What distinguishes this measure from the
ones usually encountered in particle tunneling  problems
is the fact that $\varphi$ and $z$ have finite
integration ranges. In a somewhat cavalier fashion, we extend these to $[
- \infty, \infty]$ and $[ - \infty, \infty]$ and absorb the change in
normalization by multiplying by an extra overall normalization factor $C$.
The extension of the range for $\varphi$ from $[ 0, 2 \pi]$ to $[
- \infty, \infty]$ is very natural -- it allows for motion in which the
spin direction rotates around in the same direction many times. The
justification for extending the integration range $[-r, r]$ for $z$ from
$r = 1$ to $r = \infty$ is more tenuous. The core of the
argument is the following assertion, proven in appendix~\ref{appA}:
In the presence of a $z^2$-term in the Hamiltonian, the
value of $r$ determines the degree of non-differentiability
(``wildness'') of the $\varphi$-paths that result after $z$ has been
integrated out: if $r = \infty$, the $\varphi$-paths are Brownian motion
paths; if $r=1$, they are much more ill-behaved than Brownian motion paths.
 Therefore, changing the integration range for $z$ from $r=1$ to
$r= \infty$ is equivalent to restricting attention to Brownian motion
$\varphi$-paths instead of a larger class of paths that are much more
ill-behaved. It is argued that this should not have a noticeable effect, for
the following two reasons:
if one is interested mainly in the effect of small fluctuations around
the classical path (as, for example, when calculating the tunnel
splitting energy), physically, one only expects some {\em smooth}\/
 paths in the immediate neighborhood of the classical path to be important.
Also, the additional paths that are formally included in the path integral
when the integration ranges are extended from $r=1$ to $r = \infty$
are far from the tunneling path and
therefore only make an exponentially small contribution to the path integral.

In the course of making a change of variables $(\varphi, z) \rightarrow
(\dphi, \dz)$ in the path integral in order to evaluate $\NN^{\pm}$,
the overall constant $C$ will be
multiplied by a Jacobian factor $J^N$, which may  be infinite in the
limit $N \rightarrow \infty$. To obtain finite answers, we stipulate, as
is usual, that $C$ be chosen such that the final path integral for $\dphi$
 (after the $\dz$-dependence has been integrated out), should have the
same normalization as that employed by Coleman in his discussion of
instantons~\cite{Col85}, the results of which we intend to use.
This often-used procedure may seem somewhat arbitrary, but the fact that
the path integral is defined as an infinite {\em product}\/ of integrals,
each of which may produce finite prefactors when being manipulated,
leaves one no choice but to absorb all infinities in a single
appropriately chosen constant $C$. This much having been said, we
henceforth pay no attention to normalization constants or Jacobians.

\subsection{The relative phase of the prefactors $\NN^+$ and $\NN^-$}

\label{relphase}
Before integrating out $z$, let us investigate the relative phase of
$\NN^+$ and $\NN^-$, the $(+)$-instanton and $(-)$-instanton prefactors. The
quantity $\SS - \SS_o$ that appears in the integrands of the prefactors
in eq.~(\ref{eqL3ab})
can be written in terms of the fluctuations around the tunneling path,
$(\dphi, \delz) \equiv ( \varphi - \lphi, z - \lz) $, as
\be
\label{eq125.6}
\SS - \SS_o \, = \, \int \! d \tau \left[ - i  \hbar  n s \dot{\dphi}
 \delz \, + \, a \delz^2 \, + \, b \sum_{n=2}^{\infty} {\textstyle
\frac{1}{n!} } \left.
\left(\frac{\partial^n}{\partial \varphi^n} f (\varphi)
\right) \right|_{\lphi} \dphi^n \right] \; .
\ee
Evidently, because $f (\varphi ) = f (- \varphi)$,
the following symmetry relations hold (the square brackets denote
a functional dependence, ``$\ast$'' denotes complex conjugation):
\bea
\label{eq124.9}
(\SS - \SS_o) [ \lphi^- , -\dphi, \delz] & = & (\SS - \SS_o)^{\ast} [
 \lphi^+,  \dphi, \delz ] \; , \\
\label{eq124.10}
(\SS - \SS_o) [ \lphi^- , - \dphi, - \delz] & = & (\SS - \SS_o) [ \lphi^+,
\dphi,  \delz ] \; .
\eea
Now, after the change of variables $(\varphi,z) \rightarrow (\dphi, \dz)$
in the path integral~(\ref{eqL3ab}),
$\dphi$ and $\delz$ are dummy variables that are
integrated over. Since we extended the integration ranges for $\varphi$ and
$z$ to $[ - \infty, \infty]$, the integration ranges for $\dphi$ and $\dz$
are symmetric around $\dphi = 0$ and $\dz = 0$. Hence the following
conclusions follow immediately: \newline
(i) Relation (\ref{eq124.9}) implies that $\NN^- = {\NN^+}^{\ast}$. \newline
(ii) Relation (\ref{eq124.10}) implies that $\NN^- = {\NN^+}$. \newline

As discussed in section~\ref{intz}, the extension of the integration
range to $[ - \infty, \infty]$ is on somewhat less firm ground for $z$ than
for $\varphi$. Hence conclusion (ii), which holds only if the
$\dz$-integration range
 is symmetric about $\dz = 0$, is in a sense ``weaker'' than
conclusion (i).  However, even if
the original $z$-integration range of $[-1, 1]$ is retained, the error in
the relation
$\NN^+ = {\NN^-}$ is expected to be exponentially small.
For example, for the case $z_g = - \toh$, which we shall use in
section~\ref{Kag},
only paths for which $\dz \in [ \frac{1}{2}, \frac{3}{2} ] $
break the $\dz \leftrightarrow - \dz$ symmetry, and these paths deviate so
strongly from the tunneling path (for which Re$[z] = z_g$)
 that their contribution to the path integral is exponentially small.
(Of course, this argument breaks down when $z_g = \pm 1$.)

Finally, note that within
 the lowest order of the steepest descent approximation,
in which one keeps only the $n=2$ term of the infinite sum in
eq.~(\ref{eq125.6}) (to be indicated by a superscript $(2)$ in
eq.~(\ref{eq126.2}) below), the relation
$\NN^+ = {\NN^-}$ holds, {\em independent}\/ of the range of $z$-integration.
The reason is simply that since
\be
\label{eq125.6a}
(\SS - \SS_o)^{(2)} \, = \, \int \! d \tau \left[ - i  \hbar s    \dot{\dphi}
 \delz \, + \, a \delz^2 \, + \, \toh  b  f'' (\lphi) \dphi^2 \right] \; ,
\ee
where $f'' = \partial^2 f / \partial \varphi^2$, we have
\be
\label{eq126.2}
(\SS - \SS_o)^{(2)} [ \lphi^+ , \dphi, \delz]  =  (\SS - \SS_o)^{(2)}
[ \lphi^-, \dphi, \delz ] \; .
\ee

\subsection{Calculating $\Delta$}

\label{phipath}
We henceforth  restrict attention to the lowest order of the steepest descent
approximation, in which $\NN^+ = \NN^- \equiv \NN$. It is straightforward,
starting from eq.~(\ref{eq125.6a}), to
perform the Gaussian integral over $\dz$ to arrive at
\be
\label{eq89.7}
\NN \propto \int \!  \DD \dphi e^{- \int \! d \tau \delta^2
\LL_{\varphi} / \hbar} \; ,
\ee
\be
\label{eq89.8}
{\rm where} \qquad \delta^2 \LL_{\varphi} \, =\, {\textstyle \frac{1}{2}}
b f'' (\lphi ) (\dphi)^2 +
\frac{( \hbar n s)^2}{4 a} ({\dot{\dphi}})^2 \; .
\ee
Now note that the $\delta^2 \LL_{\varphi}$ of eq.~(\ref{eq89.8}) is also
the second variation of the
following effective Euclidean Lagrangian for $\varphi$:
\be
\label{eq91.6}
\LL_{\varphi} \, = \, \toh \frac{ (\hbar n s)^2 }{2 a} \dot{\varphi}^2 \,
+ \, b \, f ( \varphi) \; .
\ee
The Lagrangian equation of motion resulting from $ \LL_{\varphi}$
is just eq.~(\ref{eq9}), and hence
the  $\SS_o$ corresponding to $\LL_{\varphi}$ will be equal to
the ${\rm Re }[ \SS_o] $ found earlier. Consequently, both $\NN$ {\em and}\/
${\rm Re } [\SS_o] $ for the original system are equal to those arising from
$\LL_{\varphi}$. It follows that $\Delta$, too, can be calculated directly
using $\LL_{\varphi}$. Furthermore,  $\LL_{\varphi}$ is
quadratic in $\dot{\dphi}$
and therefore the methods discussed by Coleman \cite{Col85} can be used to
 calculate the tunnel splitting energy. Coleman's methods, which are
summarized in appendix~\ref{appB}, readily lead to the following
expression for $\Delta$ (compare eqs.~(\ref{eq2.9}) and~(\ref{eq93.8})):
\be
\label{eq95.7}
\Delta \, = \, 2  \varphi_{\rm as} (\pi n s)^{- \frac{1}{2}}
 (b f'' (0))^{\frac{3}{4}} (2 a)^{\frac{1}{4}}
\left( e^{- \SS_o^+ / \hbar} + e^{- \SS_o^- / \hbar} \right) \; .
\ee
Here the constant $\varphi_{\rm as}$ is to be read off  from the asymptotic
behavior of the  tunneling paths, which, as Coleman shows,
can always be written in the form (compare eq.~(\ref{eq92.8}):
\be
\label{eq92.8a}
\lphi^{\pm} (\tau) \simeq \pm \pi \mp \varphi_{\rm as} e^{- \omega \tau } \; ,
\qquad{
\rm with } \; \omega \equiv \frac{ \sqrt{2 a b f'' (0)}}{\hbar n s} \; ,
\ee

Eq.~(\ref{eq95.7})
is the main result of this section. We emphasize once again that
$\Delta$ is strictly zero if
$S_o^+ - S_o^- = i \hbar l \pi$, with $l$ an odd integer, since then
 $(+)$-instanton and $(-)$-instanton events interfere destructively.

To find $\varphi_{\rm as}$ explicitly, one needs some knowledge of the
asymptotic behavior of the shape function $f(\varphi)$ of the potential
as $\varphi \rightarrow 0$. According to eqs.~(\ref{widef}),
(\ref{newf}) and (\ref{phiast}) in appendix D, the form of
$f(\varphi)$ that is applicable to the hexagon tunneling problem to
be studied in the latter parts of this paper is:
\FL
\be
\label{eq15.1}
f(\varphi)  =  \widef \left( \sqrt{ (\varphi^2_{\rm eff}
 + \past{}^2 } \right)
- \widef ( \past) \, , \qquad  \varphi_{\rm eff} \equiv \varphi \, {\rm mod}
\, \pi \in ( - \pi/2, \pi/2] \; ,
\ee
\be
\label{eq15.2}
{\rm where } \qquad  \widef ( \varphi)  =
| \sin \varphi | ( 1 + \textstyle{ \frac{1}{9}} \sin^2 \varphi
)^{\frac{1}{2}} - \wJ \sin^2 \varphi \; .
\ee
Here $\wJ \simeq 0.42$ and $\past = 0.14 \, s^{- 1/3}$.
In the limit $s \gg 1$, $\past$  can be treated as a small
parameter which characterizes the curvature of the potential at
$\varphi = 0$, since $f''(0) = (1/ \past - 2 \wJ )$.

``Energy conservation'' along the tunneling path implies for the
$\LL_{\varphi}$ of eq.~(\ref{eq91.6}) that
\be
\label{eq15.4}
\toh \frac{ ( \hbar n s )^2}{2 a} \dot{\lphi}^2 - b f ( \lphi ) = 0 \; .
\ee
Integrating this equation gives, for an $(-)$-instanton,
\be
\label{eq15.6}
\tau \omega ( f'' (0) )^{- \frac{1}{2}} \, = \, - \int_{- \pi/2}^{ - \pi +
\lphiT} \frac{ d \lphi}{[ 2 f ( \lphi ) ]^{\frac{1}{2}} } \; ,
\ee
where $-\pi + \lphiT$ is the angle reached at time $\tau$ if $\lphi = 0$
at $\tau = 0$. This equation is to be solved for $\lphiT$ as a function
of $\tau$, in the limit $\lphiT \rightarrow 0$. In particular, we take
$\lphiT \ll
\eta \past$, where $\eta \ll 1$ is an arbitrary small parameter (e.g. $\eta
= 0.01$).

It is convenient to split the integral into two parts by writing
\be
\label{eq16.1}
\tau \omega ( f'' (0) )^{- \frac{1}{2}} = F ( \pi / 2) - F (\lphiT) \; ,
\ee
\be
\label{eq18.2}
{\rm where } \qquad F(\varphi) \equiv \int_{\eta \past}^{\varphi}
\frac{ d \lphi}{[ 2 f ( \lphi ) ]^{\frac{1}{2}} } \; ,
\ee
and the property $f(\varphi) = f(\varphi + \pi)$ has been exploited.
Since $\lphiT \ll \eta \past \ll 1 $, one may
evaluate $F( \lphiT)$ analytically by using the
asymptotic form of the integrand,
namely $f(\varphi \rightarrow 0) \simeq \varphi^2 (1/2 \past
- \wJ ) $, with the result
\be
\label{eq18.6}
F(\lphiT) = \left( \frac{\past}{1 - 2 \wJ \past} \right)^{\frac{1}{2}}
( {\rm ln} \varphi \, - \, {\rm ln} \eta \past ) \; .
\ee
Using this result in eq.~(\ref{eq16.1}) and solving for $\lphiT$ gives
\bea
\label{eq19.3}
\lphiT & = & \varphi_{\rm as} e^{- \omega \tau} \; , \\
\label{eq19.4}
{\rm where} \qquad \varphi_{\rm as} &=& \exp \left[ C \left( \frac{1 - 2 \wJ
\past}{ \past} \right)^{\frac{1}{2}} \right]
\\
\label{eq18.7}
{\rm and} \qquad C & \equiv & F (\pi / 2) + \left( \frac{\past}{1 - 2 \wJ
\past} \right)^{\frac{1}{2}} {\rm ln} \eta \past
\eea
The constant $C$ is independent of $\eta$, since the
$\eta$-dependence of $F(\pi / 2)$ cancels that of the second term.
Moreover, $C$ is only very
weakly dependent on $\past$ and hence on $s$, and approaches a
constant value as
$s \rightarrow \infty$.  (Numerically it is
found that $C$ changes from 1.1 to 2.1 as $s$ changes from
1 to $\infty$.) Hence, recalling that $\past = 0.14 s^{-\frac{1}{3}}$,
 we conclude that in the limit $s \rightarrow \infty$,
\be
\label{eq19.5}
\varphi_{\rm as} \simeq e^{5.6 \, s^{\frac{1}{6}} } \; .
\ee
This expression diverges as $s \rightarrow \infty$, but in eq.~(\ref{eq95.7})
it premultiplies an expression that tends to zero even faster as $s
\rightarrow \infty$ (in section \ref{Kag}, we find $\SS_o \sim s^{1/2}$, see
eq.~(\ref{eq86.9}) and eq.~(\ref{sdep})).

\section{Spin tunneling in the Kagom\'{e} lattice}

\label{Kag}

The stage has now been set for the study of the hexagon tunneling event on the
Kagom\'{e}-lattice, for which we adopted the Hamiltonian $\hh_{\rm hex}$
defined in eqs.~(\ref{bigh}) to~(\ref{cop}).
Rewriting $\hh_{\rm AFM}$ in terms of $\varphi_l$ and $z_l \equiv
\cos \theta_l$, the Euclidean Lagrangian to be studied is
\bea
\label{eq2.3a}
\LL & = &
\sum_{l = 0}^5 - i \hbar s \dot{\varphi}_l ( z_l - 1 ) \: + \,
 J_b f( \varphi_{\rm av}) \\
\nonumber
& & + \; s^2 J \sum_{l = 0}^5 \bigl[   z_l + z_{l+1} + z_l z_{l+1} +
C_l \cos (\varphi_l -
\varphi_{l+1} )  + {\textstyle \frac{3}{2}}
\bigr] \; ,
\eea
\be
\label{eq2.6}
{\rm where} \quad  C_l \equiv \sin \theta_l \sin \theta_{l+1}
=  \{ ( 1- z_l^2) ( 1 - z_{l+1}^2 ) \}^{\frac{1}{2}} \; .
\ee
The tunneling event was described in intuitive terms in section~\ref{hte}.
The initial and final states are:
\bea
\label{eq3.2}
|i \rangle :&\quad & \varphi_l (-T/2) = (0, \pi, 0, \pi, 0, \pi ) \; ; \qquad
 z_l (- T/2) =  z_g \, . \\
|f \rangle :& \quad & \varphi_l (T/2) = (\pi, 0, \pi, 0, \pi, 0) \;  ;
 \qquad \quad z_l (T/2) =  z_g \, .
\eea
Here $z_g = \cos 2 \pi /3 =
- \toh$. The motion of each of the six spins will be
roughly analogous to the single-spin motion described in
section~\ref{tunnpath}.

\subsection{Collective coordinates}

We now introduce a coordinate transformation to a set of collective
coordinates, $(\whphi_l, \whz_l)$, which ultimately allows us to map the
hexagon tunneling problem exactly onto the simple model problem of
section~\ref{mod}, eqs.~(\ref{eq87.1}) and~(\ref{eq87.1a}):
\bea
\label{eq79.2}
z_l - z_l(-T/2) &= & \sum_{k=0}^5 e^{i \pi l k / 3} \, i \whz_k \; , \\
\label{eq79.3}
\varphi_l - \varphi_l (- T/2) & = & \sum_{k=0}^5
e^{- i \pi l k / 3}  \whphi_k \; , \\
\label{eq79.5}
{\rm where}  \qquad \whphi_l^{\ast}
=   \whphi_{-l} \; , \quad & & \whz_l^{\ast} = \whz_{-l} \; .
\eea
The conditions~(\ref{eq79.5}) ensure that $\varphi_l - \varphi_l (-T/2)$
is purely real and $z_l - z_l(-T/2)$ purely {\rm imaginary}. The reason
for the latter requirement is the same as that encountered in the
discussion of the model tunneling problem (section~\ref{mod}): it
would otherwise be impossible to satisfy $\hh = 0$.
Making $z_l - z_l(-T/2)$ imaginary flips the sign of the $z^2$-terms,
thus turning the $z$-valley into a ridge
around which the tunneling path can proceed along a constant energy
contour.

The new coordinates are essentially discrete Fourier transforms of the
original ones with respect to $l$. (Indices such as $l$ and $k$ will
be used interchangeably both as ``position'' labels (for $\varphi$ and
$z$) and as Fourier labels (for $\whphi$ and $\whz$).)
The zeroth ones, $(\whphi_o,  \whz_o)$, are just the average values
of the old coordinates and can be thought of as the coordinates of a collective
degree of freedom (of spin $6s$, as it turns out). They will be referred
to as the {\em tunneling coordinates},\/ since $\whphi_o$ tunnels from 0
at $\tau = - T/2$ to $ \pm \pi$ at $\tau = T/2$, and $i \whz_o$ is its
conjugate momentum. The other coordinates $(\whphi_l,  \whz_l)$, $l = 1,
\dots, 5$, will be called {\em transverse coordinates},\/ since they will be
shown to be strictly zero along the tunneling path, and hence describe
fluctuations that are orthogonal to the tunneling path.

In order to calculate the tunnel splitting energy
$\Delta$, one needs the classical action $\SSo{\pm}$ and the prefactor
$\NN^{\pm}$. In the following two sections, we show that
both these quantities can be obtained from a rather simple {\em effective}\/
Lagrangian,  $\wLL_o$, that depends
{\em only}\/ on the tunneling coordinates $(\whphi_o,  \whz_o)$, so that the
problem is substantially simplified. This Lagrangian, defined in
eq.(\ref{eq80.4}) and given explicitly in eq.(\ref{eq83.7}), turns out to have
just the form of the Lagrangian introduced in section~\ref{modham}, with
$(\whphi_o, \whz_o)$ corresponding to $\varphi, z)$ there.

\subsection{Classical action (Kagom\'e lattice)}

\label{So}
The classical action is completely determined by the tunneling path, for
which, as in section~\ref{mod},
 we again use overlined variables, $(\lphi_l, \lz_l)$ or $(\whlphi_l,
\whlz_l)$. This path is determined by the equations of motion, which, by
the chain rule, can be written in terms of the new coordinates as:
\bea
\label{eq81.5}
\partial_{\tau} \! \left( \frac{ \partial \LL }{ \partial
\dot{\varphi_l } } \right)_t
\left(\frac{ \partial \dot{\varphi_l}}{ \partial
\dot{\whphi_k}} \right)_t & = &
\left( \frac{ \partial \LL }{ \partial {\varphi_l } } \right)_t
\left( \frac{ \partial {\varphi_l}}{ \partial
\whphi_k} \right)_t \; ,
\\
\label{eq81.6}
\left( \frac{ \partial \LL }{ \partial {z_l } } \right)_t
\left( \frac{ \partial z_l}{ \partial  \whz_k} \right)_t
&=& 0 \; ,
\eea
with boundary conditions
$\whlphi (- T/2 ) = 0$ and $\whlz (- T/2) = 0$.
The bracket $( \;)_t$ indicates that the derivatives are to be evaluated
along the tunneling path.
These equations can be solved by the following ``tunneling path Ansatz":
\be
\label{eq81.7}
( \whlphi_l, \whlz_l ) \equiv (0, 0) \; , \quad {\rm for} \; l = 1, \dots , 5
\; .
\ee
To see that this Ansatz works, consider eqs.~(\ref{eq81.5}) and~(\ref{eq81.6})
 term by term:
\be
\label{eq81.8}
\partial_{\tau} \left( \frac{ \partial \LL }{ \partial
\dot{\varphi_l } } \right)_t \left( \frac{ \partial \dot{\varphi_l}}{
 \partial \dot{\whphi_k}} \right)_t \, = \,
- i \hbar s \dot{\lz}_l \: e^{- i \pi l k / 3} \; ;
\ee
\bea
\left( \frac{ \partial \LL }{ \partial {\varphi_l } } \right)_t
\left( \frac{ \partial \varphi_l}{ \partial
\whphi_k}  \right)_t & = & \left\{
s^2 J \left[ - C_l \sin ( \lphi_{l} - \lphi_{l+1} )
+ C_{l - 1} \sin  ( \lphi_{l-1} -  \lphi_l ) \right] \right.
\label{eq81.9} \\ \nonumber
&  + & \!\left. \vphantom{s^2 J \left[ - C_l \sin (
\lphi_{l} - \lphi_{l+1} )
    + C_{l-1} \sin  ( \lphi_{l-1} -  \lphi_l ) \right]}
 {\textstyle \frac{1}{6}} J_b  f' (\varphi_{\rm av})
\right\} e^{- i \pi l k / 3}  \; ;
\eea
\bea
\left( \frac{ \partial \LL }{ \partial {z_l } } \right)_t
\left( \frac{ \partial z_l}{ \partial  \whz_k} \right)_t  & = &
\left\{ s^2 J \left[ \lz_{l+1} + \lz_{l -1} + 2 +
\sum_{m=0}^5  \left( \frac{  \partial
C_m }{\partial z_l} \right)_t \cos ( \lphi_m - \lphi_{m+1} )
\right] \right. \nonumber \\
& & \left. \vphantom{
\left( \sum_{m=0}^5 \frac{ \partial
     C_m }{\partial z_l} \right) }  \qquad \qquad
- i \hbar s \dot{ \lphi_l} \right\} i e^{i \pi l k / 3} \; ;
\label{eq81.10}
\eea
with
\be
\label{eq81.11}
\left( \frac{ \partial C_m }{\partial z_l} \right)_t \, = \, - \frac{1}{
C_m } \left[ \lz_m ( 1 - \lz_{m+1}^2 ) \delta_{m l} +
\lz_{m+1} ( 1 - \lz_{m}^2 ) \delta_{m+1\, l} \right] \; .
\ee
Now express $(\lphi_l, \lz_l)$, wherever they occur, in terms of
$(\whlphi_l, \whlz_l)$. Using the tunneling path Ansatz~(\ref{eq81.7})
repeatedly, it readily follows that
\be
\label{eq82.7}
\lz_l \, = \, i \whlz_o + z_g \equiv \llz \; ;
\qquad \lphi_l = \whlphi_o + \varphi_o ( - T/2) \; ;
\ee
\bea
{\rm whence}\qquad
\label{eq82.7a}
\varphi_{\rm av}  = \whlphi_o \; , \quad & & \qquad
\lphi_l - \lphi_{l+1}  = \pi \; , \\
\sum_{m=o}^5 \left( \frac{\partial C_m }{ \partial z_l} \right)_t
&=& - 2 \, \llz \; .
\eea
This means that eqs.~(\ref{eq81.8}) to~(\ref{eq81.10}) depend on
the index $l$ {\em only}\/ through the factors of $e^{\pm i \pi l k / 3}$.
Since $\sum_{l = 0}^5 e^{\pm i \pi l k / 3} = 6 \delta_{k0}$, these
equations simplify considerably to become
\bea
\label{eq82.8}
-i \hbar s  \, i \dot{ \whlz_o} \, \delta_{0k} & = &
{\textstyle \frac{1}{6}} J_b f'( \whlphi_o)  \, \delta_{0k} \; , \\
i \hbar s \, \dot{ \whlphi_o} \, \delta_{0k} & = &
 4 s^2 J i \whlz_o  \, \delta_{0k}  \; .
\label{eq82.9}
\eea
where $z_g = - \toh $ has been used. Evidently, the equations of motion
for the transverse coordinates $(k \neq 0)$ trivially reduce to zero, and
{\em only}\/ the tunneling coordinates $\whlphi_o$ and
$\whlz_o$ have non-trivial behavior.

This allows us to introduce a considerable simplification:
Write $\wLL$ for the function that results when $\LL$ is expressed in
terms of the new coordinates:
\be
\label{eq80.5}
\wLL (  \whphi_o, \whz_o ;\: \dots \: ; \whphi_5,
 \whz_5) \, \equiv \, \LL \left( \{ \varphi_k( \{
\whphi_l \} ), \: z_k ( \{ \whz_l \} ) \right) \; .
\ee
Define an {\em effective}\/ Lagrangian $\wLL_o$,
 that depends only on the tunneling coordinates, by the relation
\be
\label{eq80.4}
\wLL_o ( \whphi_o , \whz_o ) \, \equiv \, \wLL ( \whphi_o, \whz_o ; 0,0 ;
\dots , 0;0) \; .
\ee
Now, the behavior of the tunneling coordinates $\whlphi_o$ and
$\whlz_o$ can be found directly via  the effective Lagrangian $\wLL_o$,
instead of the full $\wLL$.
This follows because the equations of motion for
$\wLL_o$ are exactly eqs.~(\ref{eq82.8})
and~(\ref{eq82.9}), essentially by definition, since $ \left(
\frac{\partial \, \wLL  }{\partial \whz_o} \right)_t $, evaluated according
to the tunneling path Ansatz eq.~(\ref{eq81.7}), is {\em identically}\/
equal to
$ \left( \frac{\partial  \, \wLL_o }{\partial \whz_o} \right)_t $, with
similar comments holding for $ \left( \frac{\partial \, \wLL_o }{\partial
(\partial_{\tau} \whphi_o) } \right)_t$
 and $ \left( \frac{\partial \, \wLL_o }{\partial
\whphi_o} \right)_t$.
Furthermore, since $\SS_o$ depends solely on the tunneling path, it too can
 be calculated directly from $\wLL_o$. The effective Lagrangian $\wLL_o$
can be found from eq.~(\ref{eq80.4}) to be:
\be
\label{eq83.7}
\wLL_o (\whphi_o , \whz_o) = - i \hbar 6s (\dot{ \whphi_o} )
(\Zc - 1) + 12 s^2 J (\Zc - z_g)^2 +  J_b f( \whphi_o) \, .
\ee
It is from this expression that $\hh_{\rm eff}$ of eq.~(\ref{modelham})
was obtained. Evidently $\wLL_o$
 has just the form of the Lagrangian of eq.~(\ref{eq87.1}), with
\be
\label{eq83.8}
n = 6 \, ; \qquad a = 12 s^2 J \, ; \qquad b =  J_b \, ;
 \qquad z_g = -
\toh \; .
\ee
Consequently, all the results from section~\ref{mod} are applicable.
The classical action, from eq.~(\ref{eq8}), is
\be
\label{eq83.9}
\SS_o^{\pm} = \pm 9 i \hbar s \pi + \hbar s \sqrt{\frac{3 J_b}{J s^2}}
\int_0^{\pi} \!\! d \varphi \sqrt{ f (\lphi^+ )} \; .
\ee

{}From the above expression for the classical action one can immediately
read off the most striking result of this paper:
\be
\label{phasediff}
 e^{- \SS_o^+ / \hbar} + e^{- \SS_o^- / \hbar}\;  =  \;
\left\{ \begin{array}{l}
- 2 \exp\left[ {-  \sqrt{3 J_b/  J } \int_0^{\pi} \!\! d \varphi \sqrt{
f (\lphi^+ )} } \right] \; \quad {\rm if} \; s = {\rm integer} \; , \\
0 \; \qquad \qquad \qquad {\rm if} \; s = {\mbox{\rm half-odd-integer}} \; .
\end{array} \right.
\ee
Thus, if $s$ is half-odd-integer, the $(+)$-instanton and $(-)$-instanton
amplitudes interfere destructively, and the total tunneling
amplitude is zero (for a discussion and extension of this result, see
section~\ref{desint}).

\subsection{The prefactor (Kagom\'e lattice)}

\label{Del}
The calculation of the prefactor involves the evaluation of the path
integral
\be
\label{eq84.2}
\int \! \prod_{l=0}^{5}  \DD \Omega_l
 e^{- \int \! d \tau \delta^2 \! \LL / \hbar} \;.
\ee
where the second variation of the action is
\be \begin{array}{rl}
\nonumber
\delta^2 \LL & = \: \toh
\left( \frac{ \partial^2 \LL}{\partial z_m  \partial z_n}
 \right)_t \dz_m \dz_n  \, + \,
\left( \frac{ \partial^2 \LL}{\partial \dot{\varphi}_m \partial
z_n }  \right)_t \delta \dot{\varphi}_m \dz_n  \\
\label{eq11.4}
+ &
\left( \frac{ \partial^2 \LL}{\partial {\varphi_m} \partial
z_n }  \right)_t  \dphi_m \dz_n
\, + \, \toh
\left( \frac{ \partial^2 \LL}{\partial \varphi_m  \partial
\varphi_n}  \right)_t \dphi_m \dphi_n \; .
\end{array}
\ee
The deviations around the tunneling path, defined by
\bea
\label{eq84.4}
\delta z_l &\equiv & z_l - \lz_l \: \equiv \: e^{i \pi l k / 3}
 \whdz_k \; , \\
\label{eq84.4a}
\delta \varphi_l &\equiv & \varphi_l - \lphi_l \:  \equiv \: \;
e^{- i \pi l k / 3} \whdphi_k \; ,
\eea
are taken to be purely {\rm real}\/ (it is only the tunneling path itself
that has to become complex to minimize the action). Thus, their Fourier
transforms obey the conditions $\whdphi_l^{\ast} = \whdphi_{-l}$ and
$\whdz_l^{\ast} = \whdz_{-l}$.

Now, it can be verified that
\bea
\label{eq11.7}
\left( \frac{ \partial^2 \LL}{\partial \dot{\varphi_m} \partial
z_n }  \right)_t & = & - i \hbar s \delta_{m n} \; ; \qquad \qquad
\left( \frac{ \partial^2 \LL}{\partial {\varphi_m} \partial
z_n }  \right)_t \, = \, 0 \; , \\
\label{eq11.6}
\toh \left( \frac{ \partial^2 \LL}{\partial z_m  \partial z_n}
 \right)_t & = &  \!
\toh J s^2 \! \left\{ 2 \delta_{mn}  {\textstyle \frac{ 1 }{1 -
\llz^2}}  + ( \delta_{mn+1} + \delta_{m+1n})
(1- \! {\textstyle \frac{ \llz^2}{1 - \llz^2}} ) \right\}  \\
\toh \left( \frac{ \partial^2 \LL}{\partial \varphi_m  \partial
\varphi_n}  \right)_t  & = &
 \label{eq11.9} \toh {\textstyle \frac{1}{36}} J_b f''(\whlphi_o)
+ \toh s^2 J ( 1 - \llz^2) [ 2 \delta_{mn}  - \delta_{m+1n}
 - \delta_{m n+1} ] \; .
\eea

Obtaining the actual numerical coefficients in eqs.~(\ref{eq11.6})
and~(\ref{eq11.9}) is somewhat tedious; however,
the fact that the $m,n$-dependence enters only via $\delta$-functions of the
form $\delta_{m \, n+k}$,
which is all that is needed for the following argument, can
be anticipated directly from the fact that
the original $\LL$ of eqs.~(\ref{eq2.3a})
to~(\ref{eq2.6}) is ``translationally invariant'' (i.e.\ invariant under
$l \rightarrow l+1$). Now, when writing $\delta^2 \LL$ of eq.~(\ref{eq11.4})
in terms of the Fourier transform variables $(\whdphi_l, \whdz_l)$ of
eqs.~(\ref{eq84.4}) and~(\ref{eq84.4a}), one may utilize the relation
\be
\label{eq84.8}
\sum_{ mn =0}^5 \delta_{m \, n+k} \dz_m \dz_n  \, = \,
6 \sum_{l=0}^5 e^{i \pi l k / 3} \whdz_l \whdz_{-l} \; ,
\ee
which is essentially a consequence of the convolution theorem for discrete
Fourier transformations. Using eqs.~(\ref{eq11.7}) to~(\ref{eq84.8}) in
eq.~(\ref{eq11.4}), it follows directly
that the variables $(\whdphi_o, \whdz_o)$ decouple completely from the
others in $\delta^2 \LL$ \cite{diagonalize}.
 Hence the path integral~(\ref{eq84.2}) factorizes into a product, say
$I_1 I_2$, of two independent path integrals. $I_1$
 depends only on $(\whdphi_o, \whdz_o)$, the variations of the tunneling
coordinates around their tunneling path values:
\be
\label{eq85.4}
I_1 = \int \! \DD ( \delta  \Omega_o) e^{-  \int \! d
\tau \delta^2 \wLL_o / \hbar } \; .
\ee
In writing this equation, another shortcut has been employed:
$\delta^2 \wLL_o$ is the second variation of the simple Lagrangian
$\wLL_o$ of eq.~(\ref{eq80.4}), namely
\be
\label{d2LL}
\delta^2 \wLL_o = - i \hbar 6 s \, \dot{\whdphi_o} \whdz_o \, + \, 12 s^2 J \,
\whdz^2 \, + \, \toh J_b f''( \whlphi_o ) \, \whdphi_o^2 \, .
\ee
 This simple form may be used since
$\left( \frac{\partial^2 \wLL}{\partial^2 \whz_o} \right)_t$,
evaluated according to the
tunneling path Ansatz~(\ref{eq81.7}), is {\em identically}\/ equal to
$\left( \frac{\partial^2 \wLL_o}{\partial^2 \whz_o }\right)_t$,
with similar comments
holding for the other second derivatives of $\wLL_o$ with respect to
the $\whphi_o$ and $\whz_o$.

The other path integral, $I_2$, depends only on
$(\whdphi_l, \whdz_l)_{l \neq 0}$, the variations
 of the {\rm transverse}\/ coordinates around their zero
values along the tunneling path. These fluctuations are
exactly harmonic oscillators
in the limit $\frac{J_b}{s^2 J} \rightarrow 0$, since then all time
dependence drops out of the coefficients in eqs.~(\ref{eq11.6})
to~(\ref{eq11.9}) (recall that $\llz = i \whlz_o + z_g$, with
 $i \whdz_o \propto \sqrt{b/a} \propto
\sqrt{J_b/ s^2 J}$ ). More generally, the fluctuations of the transverse
coordinates will shift
 the ground state energy by an amount proportional to
their characteristic frequencies. However, they will {\em not}\/ make a
contribution to the tunnel splitting energy $\Delta$.

Thus, for the calculation of $\Delta$, only
the path integral $I_1$ is relevant. Since this depends only on $\wLL_o$,
the demonstration that only $\wLL_o$ is important for the calculation of
$\Delta$ is complete.

It follows that $\Delta$ may be directly obtained from the effective
Lagrangian  $\wLL_o$ of section~\ref{tunnelsplit},  eq.~(\ref{eq83.7}),
inserted into the results of section~\ref{phipath}.
 Substituting the values of eq.~(\ref{eq83.8})
and~(\ref{eq83.9}) into eq.~(\ref{eq95.7}) gives
\FL
\be
\label{eq86.9}
\Delta  = \left\{ \begin{array}{l}
4 \left( {\textstyle \frac{2}{3}}\right)^{\frac{1}{4}}
\pi^{- \frac{1}{2}} \varphi_{\rm as} ( J_b
f'' (0) )^{\frac{3}{4}}
J^{\frac{1}{4}} \exp \left[ -   \sqrt{ 3 J_b / J }
\int_0^{\pi} \!\! d \varphi \sqrt{
f (\varphi )} \right]   \; \quad {\rm if} \; s = {\rm integer} \; , \\
\quad 0 \; \qquad \quad \qquad
\qquad \qquad \qquad {\rm if} \; s = {\mbox{\rm half-odd-integer}} \; .
\end{array} \right.
\ee

This is the main result of this paper -- an explicit expression for the
tunnel splitting energy in a non-trivial setting.

Expression (\ref{eq86.9})
can be evaluated numerically (see the last paragraph of
appendix~\ref{appD} for the definition of $J_b$ and $f(\varphi)$, and
eqs.~(\ref{eq19.4}), (\ref{eq18.7}) and (\ref{eq18.6}) for
$\varphi_{\rm as}$).
The results for $\Delta$, in units of $J$, for
some values of $s$ are tabulated in table~\ref{tab1}.
 It should be remarked, though,
that strictly speaking and for several reasons,
our methods are only applicable in the limit of
$s \gg 1$. Collecting the $s$-dependencies of $\varphi_{\rm as}$,
$f''(0)$ and $J_b$ gives
\be
\label{sdep}
\Delta  \propto s J \exp [ 5.6 s^{1/6 } - 5.6 s^{1/2} ] \qquad {\rm for}
\; s \gg 1 \; .
\ee

\begin{tabular}{|c||c|c|c|c|c|c|c|c|} \hline
$s$ & 1 & 2 & 3 &   4 & 5 & 6 & 10 & 20 \\
$\Delta/ J$ & 2.3 & 0.83 & 0.32 & 0.13 & 0.06 & 0.02 & 0.001 & 5
$\times 10^{-6}$ \label{tab1} \\ \hline
\end{tabular}

\section{Destructive interference}

\label{desint}
It is quite striking
 that the tunnel splitting energy and the tunneling amplitude (see
eqs.~(\ref{phasediff}) and~(\ref{eq86.9})) are
exactly zero if $s$ is a half-odd-integer. This is a consequence
of destructive interference of the $(+)$-instanton
and $(-)$-instanton amplitudes
in eq.~(\ref{phasediff}). Actually, this result can be arrived at without the
need for a detailed calculation (see also \cite{vDH92a}).
All that is needed is the fact that
$Re [ \SS_o^+ ] = Re [ \SS_o^- ]$ and $\NN^+ = \NN^-$, which is guaranteed
by symmetry (under some technical assumptions, explained in
section~\ref{relphase}), and knowledge of
 the phase of the classical action,
eq.~(\ref{eq83.9}). In fact, a
similar result holds for a much larger class of tunneling problems on the
Kagom\'{e} lattice.

Consider the simultaneous tunneling of
larger sets of spins on the Kagom\'{e} lattice. Take, for example, any
closed ``loop" of spins such that all the spins on the loop alternate
between the type A and type B around the loop.
It is proven in appendix~\ref{appC} that all
 such loops contain $4n + 2$ spins (with $n$ some integer).
 Now consider a ``generalized weathervane mode'',
i.e.\  tunneling between configurations $| i \rangle$
  and $|f \rangle$
that differ from each other only through $\phi_l \rightarrow \phi_l + \pi$
for each spin on the loop (i.e. in that all A- and B-spins
on the loop are interchanged).
This generalized weathervane mode does not cost any classical exchange
energy, but selection effects provide a coplanarity barrier that has to
be tunneled through. Under the (admittedly somewhat tenuous)
assumption that during the tunneling motion of the $4n + 2$ spins all
other spins will remain fixed, the following assertions can be made:
 The absolute values of the tunneling amplitudes of the $(+)$-instanton and
$(-)$-instanton events are equal, and the relative phase between them is
 $i 2 \pi s (4n +2) (- \frac{3}{2})$. Hence, for half-odd-integer $s$,
destructive interference
between $(+)$-instantons and $(-)$-instantons occurs again.

Only when two loops, with a total number of $4n$ spins,
 tunnel simultaneously and synchronously will there
be constructive interference between all the $4n$ spins. The smallest
group of spins for which this is conceivable is the 6-spin inner loop of
a hexagon, nested inside a larger loop of 18 spins. These two loops
would tend to tunnel synchronously, because of the coplanarity forces
between the triangles connecting them.
Thus, the smallest tunneling event that has a non-zero amplitude involves
a double loop of altogether 24 spins.

The fact that destructive interference happens
for half-odd-integer $s$ is noteworthy, since $s$ is half-odd-integer for
the two experimental realizations of the Kagom\'{e} lattice that have
been proposed. These are the  magnetoplumbite $SrCr_{8-x}Ga_{4+x}O_{19}$
($s= \frac{3}{2}$) \cite{OLI88,REC90,BAE90}
 and the second layer of $He^3$ atoms on a graphite substrate ($s=
\frac{1}{2}$) \cite{Gre90,Els90}.

\section{Conclusion}

\label{discussion}

\subsection{Discussion: Comparison of energy scales}

In this paper, the tunneling barrier was taken as a fixed potential,
given by
$\hh_{\rm cop}$ in eq.~(\ref{cop}), although the original Hamiltonian
(\ref{hhexp}) had {\em no}\/ ``coplanarity forces''. In what follows, we
argue that this approach can be justified in the limit of large $s$.

Our estimate of $\hh_{\rm cop}$ in appendix~\ref{appD} is based on the
following strategy: The tunneling hexagon is frozen at a given point along
its tunneling path, at which its plane is tilted by $\varphi$ relative to
the reference plane. Then the frequencies $\omega_l ( \varphi)$ of the
zero-point modes of the spins on neighboring hexagons are calculated
as a function of $\varphi$. The coplanarity barrier is then proportional to
$\sum_l \toh \hbar ( \omega_l ( \varphi) - \omega_l (0) )$.

This procedure is reasonable if the tunneling process is ``adiabatically''
slow, i.e. if the duration of an instanton, say $\tau_{\rm in}$, is indeed
long compared to the typical oscillation periods $\omega_l^{{}^{-1}} (\varphi)$
 of the zero-point modes on the surrounding hexagons. In this case, we are
effectively ``integrating out'' in a crude way the fast modes of a
complicated many-spin problem. However, if for some zero-point modes one
has $\omega_l^{{}^{-1}} (\varphi) \stackrel{>}{\sim} \tau_{\rm in}$, these
modes are slower than the instanton and do not have time to affect the
barrier. In that case, the effective barrier should be smaller than the one
we used, since the contributions of slow modes to $\hh_{\rm cop}$ should not
be counted.

Now, in appendix~\ref{appD}, the behavior of the zero-point modes of the
hexagons surrounding the tunneling hexagon is found to be as follows: There
are both ordinary and soft modes (corresponding to $\rho_{l \neq 3}^{\alpha}$
and $\rho_{l=3}^{\alpha}$, respectively, in appendix~\ref{appD}).
For the ordinary modes, we find $\hbar \omega_l(\varphi) \sim s J$, as is
usual for antiferromagnets. The soft modes are weathervane-like
fluctuations; as $\varphi \simeq 0$, we find $ \hbar \omega_{\rm soft}
(\varphi)
\sim s^{\frac{2}{3}} J $, while for large $\varphi$, one has
$\hbar \omega_{\rm soft} (\varphi) \sim s J$.
If one were to consider fluctuations on the entire lattice (instead of only
on the neighboring hexagons of the tunneling hexagon), these soft modes
would translate to an entire brach of soft modes, for which one would also
expect $ \hbar \omega_{\rm soft}  \sim s^{\frac{2}{3}} J$.
This, indeed, is the result found by Chubukov \cite{Chu92}, using a method
based in Fourier space.

The duration of an instanton event, on the other hand, scales with $s$
according to $\tau_{\rm in} \sim 1/d
\sim \hbar/ (J s^{\frac{1}{2}})$, according to eqs.~(\ref{eq9})
and~(\ref{eq83.8}). Thus, in the limit of $s \gg 1$, one has $\tau_{\rm
in} \gg \omega_l^{{}^{-1}}$ for all zero-point modes,
both ordinary and soft. In this limit, the instanton is thus
indeed slow compared to the zero-point fluctuations, justifying the use of a
coplanarity barrier $\hh_{\rm cop}$ for sufficiently large $s$.

How does the calculated tunnel splitting energy $\Delta$ compare to the
in-plane selection energy (of order $J_c$) mentioned in
section~\ref{selectioneffects}, that tends to lift the degeneracy between
$\iii$ and $\fff$, and hence to suppress tunneling? According to
Chubukov \cite{Chu92} (see also \cite{Sac92}),
$J_c$ is a small fraction of the zero-point soft-mode energy,
i.e. $J_c \sim \eta J s^{2/3}$, with $\eta < 0.05$. Since we found
$J_b = 5/2 Js$ (see eq.~(\ref{hcopf})), neglecting $J_c$ relative to $J_b$ is
justifiable, particularly when $s \gg 1$.
This is the justification for not adding an in-plane selection term
to $\hh_{\rm hex}$ in eq.~(\ref{bigh}) in the same way that we added
the coplanarity selection term $\hh_{\rm cop}$. However, it should be
noted that for small tunneling angles $\varphi$, for which $\hh_{\rm cop}
\sim J s^{2/3}  $, the inclusion of in-plane selection effects,
also of order $J s^{2/3}$, would
probably affect some of our results. For example, our calculation of
$\varphi_{\rm as}$ in section \ref{phipath} would be affected if
$J s \past \ll J_c \ll J_b$.

Finally, from the table on p.\pageref{tab1}, it follows that for
physically realizable (i.e.\ small) spin values, we have
$\Delta \stackrel{>}{\sim} J_c$. Thus, to the extent that our
calculations can be trusted
for small $s$, tunneling effects are at least as important as in-plane
selection in determining the true nature of the ground state. We conclude,
therefore, that for small (integer) $s$, tunneling should be regarded as a
significant disordering mechanism competing with order-from-disorder
selection effects.

In a quantum Potts model of the entire Kagome system,
the Hamiltonian  would have matrix elements proportional to
$\Delta$ between any two pairs
of states which differed by the exchange $A \leftrightarrow B$ along a single
6-spin hexagon loop (or more generally, there would be matrix
elements for any larger closed loop of
alternating AB-spins). Then a superposition
of Potts states, which takes advantage of the resonance between
different hexagons (or larger loops),
would have a lower energy (at small $s$) than the
$\sqiii$ state which is favored by the $J_c$ term.

Two possible kinds of ground states suggest themselves:
The first is a ``totally disordered'' Potts state; it
can be idealized by a trial wavefunction which is an equal
superposition of all the Potts groundstates,
which is known to be disordered (it has only algebraically
decaying  correlations \cite{HR92}).
The second kind of ground state is
a ``partially disordered'' $\sqiii$ state;
an idealized trial wavefunction has, say, all C-spins
fixed, as in the usual $\sqiii$ state, but
each ABABAB-hexagon is replaced by an equal superposition
of the ABABAB- and BABABA-configurations,
(with no correlation between one hexagon and another.)
Such a state would still have long-range order with respect to the C-spins,
but would be disordered with respect to the A- and B-spins.
We call this a ``ferrimagnetic'' ground state, since the expectation
value of $\bon_z$ is +1 on the C-sublattice, and $-\toh$ on the other two
sublattices.

Both these candidate states have considerably less order than
the long-range-ordered $\sqiii$ state.
On the other hand,
they are more ordered than the ``spin-nematic'' state
which has been previously proposed for this system
\cite{CHS92,CC91}).
In the ``spin nematic'', all spins
lie in the same plane, but all directions in a given plane are equally
likely, i.e. $\langle
e^{i \theta (\bor)} \rangle = 0$.
The states described by quantum Potts models,
have either an A-, B- or C-spin at each site,
so that $\langle e^{i 3 \theta (\bor)} \rangle \neq 0$ \cite{ABR90}.

For half-odd-integer $s$, of course, hexagon tunneling is
absent   (the smallest tunneling
event with non-zero amplitude involves the synchronous turning of
24 spins), as discussed in section~\ref{desint}. Therefore, one would
expect the  long range order of the usual $\sqiii$ state to persist,
at least to much smaller values of $s$ than for integer spins.

\subsection{Possible extension and other applications}

\label{conclusion}

A more fundamental approach for treating coplanarity selection effects,
that should be trustworthy also for small $s$,
would be as follows: write down a functional integral for the entire $N$-spin
system, starting from the Heisenberg Hamiltonian,
eq.~(\ref{hhexp}), without an explicit coplanarity term such as
eq.~(\ref{cop}).  Adopt a
tunneling path in which only the six spins of the tunneling hexagon move.
The tunneling coordinates would be essentially the same as the collective
coordinates $(\whphi_o, \whz_o)$ introduced in section~\ref{Kag}. However, one
would then have of order $N-1$ transverse modes, instead of only 5.
Integrating these out would result in an effective potential for $(\whphi_o,
\whz_o)$, which could, in principle, be non-local in time (since it encodes
the effect of changing $(\whphi_o, \whz_o)$ at time $\tau$ on the transverse
modes, which, if these modes are slow, will in turn influence the potential
felt by $(\whphi_o, \whz_o)$ at later times).

We briefly mention a possible application of tunneling calculations in
the context of finite-size systems.
Consider some frustrated antiferromagnetic system whose ground states have
spin order with non-trivial discrete symmetry.
For small system sizes, the eigenvalue spectrum can be found by
exact diagonalization of the Hamiltonian.
Tunneling calculations could be useful in
obtaining better interpretations of such an eigenvalue spectrum.
For example, the $J_1-J_2$ square antiferromagnet (for $J_2 > 0.5 J_1$)
has classical ground states  in which
the even sublattice has alternating spins of one staggered magnetization
and the odd sublattice has an independent, arbitrary staggered magnetization.
When quantum fluctuations are taken into account,
spin-wave theory gives an effective ``collinearity'' force
(analogous to the coplanarity term in eq.~(\ref{cop})
in the present paper). This
lines up the staggered magnetizations along the same axis,
in either sense \cite{Hen89,CCL89}, so that classically
there are two ground states related by a discrete
symmetry and separated by a barrier. Consequently,
 the two lowest eigenstates of a large system should both be spin singlets,
namely the symmetric
and antisymmetric combinations of the two classical ground states
(which are also spin singlets) that have been mixed by quantum tunneling.
(This behavior contrasts to that of  ordinary antiferromagnets; there the
smallest gap is associated with $E \simeq \frac{s (s +1)}{2 \chi}$, where
$\chi$ is the susceptibility, so
that the first excited state is a triplet.)
The classical path expected (in a small system) has all spins of
a given sublattice rotating together during the tunneling event,
so it should be possible to map the $J_1-J_2$ tunneling problem
to a four-spin problem,
in the same fashion that we mapped a six-spin problem to a
one-spin problem.

\subsection{Summary}

We have presented a detailed calculation of the tunnel splitting energy
$\Delta$ that arises because of the tunneling of six hexagon spins
between two degenerate ground state configurations on a Kagom\'{e} lattice.
This was done by introducing a set of collective coordinates and thus
mapping the problem onto a much simpler model problem
that can be treated by standard methods. These involve the method of
steepest descent to evaluate a coherent-spin-state path integral in
imaginary time and integrating out the $\cos \theta$-degree of freedom to
obtain an effective Lagrangian $\LL_{\varphi}$, involving only the
coordinate $\varphi$. The tunnel splitting energy for $\LL_{\varphi}$ can
then be found by using well-known
methods developed for the tunneling of particles through a double-well
barrier.

The most striking aspect of our final result, eq.~(\ref{eq86.9}), is that
 when $s$ is half-odd-integer,
the tunneling amplitude and the tunnel splitting energy $\Delta$ are
strictly zero, due to destructive interference between $(+)$-instanton and
$(-)$-instanton paths. This destructive interference is argued to occur
also for larger loops of AB-spins tunneling between two degenerate
configurations. For integer $s$, we find that the tunnel splitting energy
$\Delta$ is at least of the same order of magnitude
as the in-plane selection energy $J_c$ found by other
authors; this implies that tunneling constitutes an important mechanism
competing with order-from-disorder selection effects, which might tend
 to drive the system into a partially disordered ground state.

\acknowledgments

Helpful discussions with A. Auerbach, E.P. Chan, D.P.
DiVincenzo, V. Elser, A. Garg, M. Kvale and  S. Sachdev
are gratefully acknowledged. Related preliminary calculations were done
by Q. Sheng.  Part of this work was supported by N.S.F. grant DMR--9045787.

\appendix{}

\label{appD}

In this appendix we obtain an estimate for the form of the shape-function $
f(\varphi)$ and the magnitude $J_b$ of the coplanarity potential $\hh_{\rm
cop}$ introduced in section~\ref{hte}, eq.~(\ref{cop}).

\subsection{Estimate of zero-point energy}

In principle, $\hh_{\rm cop}$ has to be calculated as follows: Start from a
$\sqiii$ coplanar ground state, and for a given hexagon, say P,
turn all six spins
(of the types A and B, say), around the C-direction
(which we take to define the $\boz$-axis), by an angle $\varphi$.
This weathervane rotation has no cost in classical energy. However, if
one allows all spins in the lattice to execute zero-point fluctuations around
their classical ground state directions (characterized by a set of unit
vectors $\{ \bon_i^{(o)} \}$), and calculates the associated
zero-point energy, $E_o(\varphi)$, it is found that $E_o(\varphi)$ has
its minimum at
$\varphi = 0$ \cite{RCC92,Sac92,CHS92,CH92,Chu92}. Hence $E_o(\varphi)$
acts as a barrier that opposes weathervane rotations.

To obtain a crude estimate of $E_o(\varphi)$, we shall only
calculate the effect
of turning P on the zero-point energy of {\em one}
of its nearest-neighbor hexagons,
say Q (see figure~\ref{ABABAB}), arguing that the effects of
turning P should have an increasingly weak effect on more distant hexagons.
Let the six spins of Q be labelled by $l = 0, \dots, 5$ (where the
index $l$ is defined modulo 6). Thus, we allow these six spins (including the
$l = 0$ spins, which is shared by Q and P), to fluctuate,
\be
\label{D1.6}
\bon_l = \bon_l^{(o)} + \sum_{\alpha = \supx, \supy, \supz }
\sigma_l^{\alpha} \boe_l^{\alpha} \; ,
\ee
while {\em all}\/ other spins in the lattice are kept fixed at their
respective $\bon_i^{(o)}$. Here we have introduced a (righthanded)
set of local basis
 vectors, $\boe_l^{\alpha}$, ($\alpha = {\rm x, y, z }$) for each spin,
with $\boe_l^{\supz} \equiv \bon_l^{(o)}$. We consider only {\em
small}\/
deviation amplitudes $\sigma_l^{\alpha} \ll 1$, and require that
$|\bon_l| = 1$.
At the end of the calculation, we shall estimate the
effects of allowing all six
nearest-neighbor hexagons of P to fluctuate simultaneously.

Given eq.~(\ref{D1.6}),
perform the standard expansion of the  classical Hamiltonian~(\ref{hhexp})
around $\bon_l^{(o)}$, to quadratic order in $\sigma_l^{\alpha}$:
\bea
\label{D3.1}
\hh - \hh^{(o)} &=& s^2 J \sum_{\langle i, j \rangle} \left[ \bon_i \cdot
\bon_j - \bon_i^{(o)} \cdot \bon_j^{(o)} \right] \\
\label{D3.2}
& =& s^2 J \sum_{l = 0}^5 \left[ \sigma_l^{\supx}{}^2 + \sigma_l^{\supy}{}^2
+ \sum_{\alpha, \beta \in \{\supx, \supy \} } \sigma_l^{\alpha}
\sigma_{l+1}^{\beta} \boe_l^{\alpha} \cdot \boe_{l+1}^{\beta} \right] \; .
\eea
No linear terms occur, because the $\bon_i^{(o)}$-directions minimize $\hh$;
this is true even when P has been turned relative to Q, since weathervane
twists do not cost any classical energy.

Now, for each $l$, the variable $(-\hbar \sigma_l^{\supx})$
 plays the role of a canonically conjugate momentum to the variable
 $s \sigma_l^{\supy}$. Suppose that it is possible to find a
transformation to a new set of pairs of canonically conjugate variables,
$\{ s \lrho_l^{\supy}, \hbar \lrho_l^{\supx} \}$,
in terms of which $\hh - \hh^{(o)}$ is
diagonal:
\be
\label{D5.5}
\hh - \hh^{(0)} =
J \sum_{l = 0}^5 \left[ \toh \frac{s^2 (\hbar \, \lrho_l^{\supx})^2}{m_l}
+ \toh k_l (s \lrho_l^{\supy})^2 \right] \; ,
\ee
This is just another way of writing the usual spin wave theory --- the
$\lrho_l^{\alpha})$ are magnons.  The
zero-point energy of the system can then be written simply as
\be
\label{D5.4}
E_o (\varphi) \equiv \sum_{l = 0}^5 \toh \hbar [ \omega_l (\varphi) -
\omega_l (0) ] \; ,
\ee
where the frequencies $\omega_l $ and r.m.s. amplitudes of $s
\lrho_l^{\supy}$ are given respectively by
\be
\label{D5.6}
\omega_l  = s \sqrt{k_l / m_l} \qquad {\rm and} \qquad
\langle (s \lrho_l^{\supy} )^2 \rangle^{\frac{1}{2}}
= \left( \frac{\hbar}{2 m_l \omega_l} \right)^{\frac{1}{2}}  \; ,
\ee

In attempting to transform eq.~(\ref{D3.2}) into the form eq.~(\ref{D5.5}),
we proceed as follows: Make the transformations
\be
\label{D3.8}
- \sigma_l^{\supx} \equiv 6^{- \frac{1}{2}} \sum_{k = 0}^5 e^{- i \pi lk/3}
\rho_k^{\supx} \;, \qquad
\sigma_l^{\supy} \equiv 6^{- \frac{1}{2}} \sum_{k = 0}^5 e^{ i \pi lk/3}
\rho_k^{\supy} \; ,
\ee
where $\rho_{-k}^{\alpha} = (\rho_k^{\alpha})^{\ast}$.
Also make the following choices for the vectors $\boe_l^{\alpha}$
 (here $R(\bon, \varphi) $
denotes a rotation by an angle $\varphi$ around the  $\bon$ direction):
\be
\begin{array}{lll}
\boe_{1,3,5}^{\supx} = (1, 0, 0) ; \quad &
\boe_{1,3,5}^{\supy} = (0, 1, 0) ; \quad &
\boe_{1,3,5}^{\supz} = (1, 0, 1) ; \quad \\
\boe_{2,4}^{\supx} = \toh (-1, 0, -\sqrt{3} ) ; \quad &
\boe_{2,4}^{\supy} = (0, 1, 0 ) ; \quad &
\boe_{2,4}^{\supz} = \toh (\sqrt{3}, 0, -1 ) ;
\end{array}
\ee
\be
\begin{array}{lll}
\boe_o^{\supx} & = R(\boe_o^z, -\varphi) \, R (\boz, \varphi) \,
 \boe_{2,4}^{\supx} &=
\toh( -1 - \sin^2 \varphi, \, \cos \varphi \sin \varphi, \,
- \sqrt{3} \cos \varphi ) \\
\boe_o^{\supy} & = R(\boe_o^z, -\varphi) \,  R (\boz, \varphi) \,
\boe_{2,4}^{\supy} &= \toh ( - \cos \varphi \sin \varphi, \, 2 - \sin^2
\varphi, \, \sqrt{3} \sin \varphi ) \\
\boe_o^{\supz} & = R (\boz, \varphi) \, \boe_{2,4}^{\supz} &= \toh ( \sqrt{3}
\cos \varphi, \, \sqrt{3} \sin \varphi, \, -1 )
\end{array}
\ee
In the definition of $\boe_o^{\alpha}$, the $R(\boz, \varphi)$ rotation is
introduced to account for the weathervane rotation of hexagon P relative to
Q. The additional rotation
$R(\boe_o^{\supz}, -\varphi)$ for $\boe_o^{\supx}$ and
$\boe_o^{\supy}$ is
introduced merely for convenience -- it allows one to minimize
the effect of
$\boe^{\supx} \! \cdot \! \boe^{\supy}$ cross terms in  eq.~(\ref{D3.2}).

By straightforward calculations one readily arrives at
\be
\label{D4.5}
\hh - \hh^{(o)} = \hh_{\supx \supx} + \hh_{\supy \supy} + \hh_{\supx
\supy} \; ,
\ee
\bea
\label{D4.6}
\hh_{\supx \supx} & = & s^2 J \left[ \sum_{l = 0}^5 (1 - \toh \cos l \pi
/3) | \rho_l^{\supx} |^2 - { \textstyle \frac{1}{12}} \sin^2 \varphi
\sum_{k, l = 0}^5 \left( e^{i \pi k /3} + e^{i \pi l /3} \right)
\rho_k^{\supx} \rho_{-l}^{\supx} \right] , \\
\label{D4.7}
\hh_{\supy \supy}  & = & s^2 J \left[ \sum_{l = 0}^5 (1 + \cos l \pi
/3) | \rho_l^{\supy} |^2 - {\textstyle \frac{1}{12}} \sin^2 \varphi
\sum_{k, l = 0}^5 \left( e^{- i \pi k /3} + e^{- i \pi l /3} \right)
\rho_k^{\supy} \rho_{-l}^{\supy} \right] , \\
\label{D4.8}
\hh_{\supx \supy} & = & { \textstyle \frac{1}{12}} s^2 J \sin 2 \varphi
\sum_{k,
l = 0}^5 [\cos k \pi /3 - \cos l \pi/3] \rho_k^{\supx} \rho_{l}^{\supy}
\eea

Evidently, $\hh - \hh^{(o)}$ is not yet diagonal. However, one can obtain an
estimate of the zero-point energy $E_o (\varphi)$ by summing the frequencies
of the six individual ``pure'' modes and neglecting the coupling between
 different modes (to obtain a pure $l$-mode, take
$\rho_k^{\alpha} = 0$ for all $k$ except for $k = l$ and $k = -l$)
\cite{varwave}.  For pure modes,
$\hh_{\supx \supy} = 0$, and the frequency $\omega_l$ can be found from
eq.~(\ref{D5.6}). The smallest frequency is that of the mode
$\rho_3^{\alpha}$, which corresponds to weathervane-like fluctuations of
the spins of Q,
and costs zero energy when $\varphi = 0$. From eq.~(\ref{D4.5}),  one finds
\be
\label{weather}
[\hh - \hh(0)]_{l = 3} =
s^2 J \bigl[ (\rho_3^{\supx})^2 (\textstyle{\frac{3}{2}} +
\textstyle{\frac{1}{6}} \sin^2 \past ) \, + \, (\rho_3^{\supy})^2
\textstyle{\frac{1}{6}} \sin^2 \past \bigl]
\ee
\be
\label{D6.2}
\hbar \omega_3 = sJ | \sin \varphi | ( 1 + \textstyle{
\frac{1}{9}} \sin^2 \varphi )^{\frac{1}{2}} \; .
\ee
The frequencies for the other five pure modes have the form
\be
\label{othermodes}
\hbar \omega_l = s J \left[ ( c_{1l} + c_{2l} \sin^2 \varphi_l) (1 +
c_{3l} \sin^2 \varphi_l) \right]^{\frac{1}{2}} \; ,
\ee
where $c_{1l}$, $c_{2l}$ and $c_{3l}$ are constants that can be obtained
with some more work. It turns out that $c_{1l} \neq 0$ for $l \neq 3$,
so that, as expected, these modes are not soft. Also, we find that their
joint  contribution to $E_o (\varphi)$ (in which $\omega_l (\varphi = 0)$
is subtracted off)
 can be mimicked, to within $1 \%$ accuracy, by the expression
 $- 0.21 s J  \sin^2 \varphi$. Hence, we conclude that the total zero-point
energy of the fluctuations of hexagon Q when its neighbor, hexagon P, is turned
by $\varphi$, is $E_o (\varphi) = \toh J s \widef (\varphi)$, where
\be
\label{widef}
\widef (\varphi) \simeq   | \sin \varphi | ( 1 + \textstyle{
\frac{1}{9}} \sin^2 \varphi
)^{\frac{1}{2}} - \wJ \sin^2 \varphi  \; ,
\ee
and $\wJ = 0.42$.

Now we have to estimate the zero-point energy when {\em all}\/ six
neighboring hexagons of hexagon P are fluctuating simultaneously. To
do an extended version of the above calculation seems forbiddingly tedious.
Roughly, we estimate that the total zero-point energy
will be $\hh_{\rm cop} \simeq 5 E_o(\varphi)$.
We have two arguments, both rather crude,
for using a factor 5 and not 6: Firstly, there are 30 (not 36)
independent spins in the six hexagons surrounding P.
Secondly, there exists one particular superposition of the
six soft weathervane modes (i.e. a mode for which only
$\rho_3^{\alpha} \neq 0$) of the six neighboring hexagons
of P which results in a mode that is totally unaffected by turning hexagon
P.  (In this mode, only the six spins of hexagon P
and the 18 spins on the outer perimeter of the six neighboring hexagons are
fluctuating, at no cost in energy, while the six spins between the inner and
outer loops of fluctuating spins remain fixed.)

\subsection{Self-consistency of zero-point fluctuation amplitude}

In the above calculation, the zero-point motion of hexagon P itself about its
orientation at fixed $\varphi$ has been ignored. This is reasonable if
$\varphi$ is not small. However, if $\varphi \stackrel{<}{\sim}
\past$, where $\past \equiv \langle \varphi^2 \rangle^{\frac{1}{2}}$
is the r.m.s. amplitude of the zero-point fluctuations
 of a hexagon around the coplanar ($\varphi = 0$)
orientation, the zero-point motion of P itself can no longer be neglected.
It has the effect of rounding out the effective potential $f(\varphi)$
at the bottom of the well from a linear to a quadratic $\varphi$-dependence.
Arguing that even if $\varphi =0$ for a given hexagon, to its
neighbors it will actually seem to have
$\varphi \simeq \pm  \past$ because of its zero-point fluctuations,
we incorporate this effect into $f(\varphi)$ by using an effective
shape function defined by
\FL
\be
\label{newf}
f(\varphi)  =  \widef \left( \sqrt{ \varphi^2_{\rm eff}
 + \past{}^2 } \right)
- \widef ( \past) \, , \qquad  \varphi_{\rm eff} \equiv \varphi \, {\rm mod}
\, \pi \in ( - \pi/2, \pi/2] \; ,
\ee
where $\widef (\varphi)$ is given by eq.~(\ref{widef}). Thus the cusp
in the shape function at $\varphi = 0$ is rounded out (see
fig.~\ref{barrier}), while for large
$\varphi$, we essentiallly have $f (\varphi) \simeq  \widef (\varphi)$.

The r.m.s. amplitude $\past$ has to be determined self-consistently:
Calculate the r.m.s. value $\past$ for hexagon P, fluctuating about the
coplanar ($\varphi = 0)$ configuration, under the assumption
that all its neighbors are turned by $\varphi = \past$ from the
coplanar configuration, and solve the resulting
self-consistency condition for $\past$.
Since $\past$ is expected to be small, in this estimate we take into
account only the weathervane fluctuations on P,
characterized by $ \rho^{\alpha}_3$.
Expression~(\ref{weather}) gives the effective Hamiltonian
for the weathervane mode of P when P is coupled
to just one neighboring hexagon (turned by $\varphi$), while the other five
neighbors of P are kept fixed at $\varphi = 0$. When all six neighbors
are turned by roughly $\past$, the effective Hamiltonian for the
weathervane mode of P is therefore roughly
\be
\label{weathersix}
s^2 J \bigl[ (\lrho_3^{\supx})^2 (\textstyle{\frac{3}{2}} +
\textstyle{\frac{1}{6}} \sin^2 \past ) \, + \, (\lrho_3^{\supy})^2
6 \textstyle{\frac{1}{6}} \sin^2 \past \bigl]
\ee
(Note that only the ``stiffness'' (coefficient of $(\lrho_3^{\supy})^2$)
gets a factor of 6, not the ``mass'' (coefficient of $(\lrho_3^{\supx})^2$).)
Using expressions (\ref{D5.6})
and~(\ref{weathersix}), one therefore finds the following
self-consistency relation:
\be
\label{selfcon}
s \lrho_3^{\supy}{}^{\ast} =
\sqrt{ \frac{s}{2}} \left( \frac{3 J}{2 J \past{}^2}
\right)^{\frac{1}{4}} \; .
\ee
Tracing  the transformations from $\lrho_3^{\supy}$
back to $\varphi$, one finds that the deviation $\lrho_3^{\supy}{}^{\ast}$
corresponds to hexagon P being turned by $\past =
\frac{\sqrt{2}}{3} \lrho_3^{\supy}{}^{\ast}$. Using  this in
eq.~(\ref{selfcon}) and solving for $\past$ one readily finds
\be
\label{phiast}
\past = \textstyle{\frac{1}{9}} \sqrt{\frac{3}{2}}
 s^{-\frac{1}{3}} \, = \, 0.14 s^{-1/3} \; .
\ee
Note that $\past \rightarrow 0 $ as $s \rightarrow \infty$.
The $s^{- 1/3}$-dependence here is in agreement with the work of Chubukov
\cite{Chu92}. He has used a self-consistent theory which is similar in
structure to our arguments in this appendix,
but based in Fourier space (and taking into account cubic terms of the form
$\sigma^{\supx} (\sigma^{\supy})^2$, $\sigma^{\supx}{}^3)$,
to calculate zero-point frequencies and amplitudes of fluctuations.

The various frequencies $\omega_l (\varphi )$ scale as follows with
$s$: For the soft mode ($l = 3$), eqs.~(\ref{newf}) and~(\ref{D6.2})
show that $\hbar \omega_3 (\varphi) \sim s J  \varphi^2/ \past$
as $\varphi \rightarrow 0$, so that, by eq.~(\ref{phiast}), one has
$\hbar \omega_3  \sim s^{2/3} J$ for $\varphi \stackrel{<}{\sim} \past$.
For $\varphi \gg
\past$, one has $\hbar \omega_3 (\varphi) \sim s J$. For the five non-soft
modes $(l\neq 3)$, eq.~(\ref{othermodes}) shows that $\hbar \omega_l
(\varphi) \sim s J$, for all $\varphi$.

To summarize the results of this appendix: The coplanarity potential has the
form
\be
\label{hcopf}
\hh_{\rm cop} = 5 E_o (\varphi) = J_b f (\varphi) \, ,
\ee
with $J_b = \textstyle{\frac{5}{2}} Js $. The functions
 and $f(\varphi)$ and $\widef
(\varphi)$ are defined in eqs.~(\ref{newf}) and~(\ref{widef}) respectively,
and $f (\varphi)$ depends on $\past \sim s^{- 1/3}$. $f (\varphi)$ is plotted
in fig.~\ref{barrier} for the case $s = 1$.
Our crude estimates are expected to be reliable to within a factor of 2.

\appendix{}

\label{appA}
In this appendix, which is a complement to section~\ref{intz},
 we investigate the way in which the integration range $[-r,
r]$ for $z$ influences the degree of non-differentiability of the
$\varphi$-paths that result after $z$ has been integrated out, for a
Hamiltonian containing a $z^2$-term. It is shown
that for $r=1$,
the resulting $\varphi$-paths are highly non-differentiable, much more
pathological
than Brownian motion paths. For $r= \infty$ they are Brownian motion
paths.

To see the difference between $r=1$ and $r= \infty$, consider
the simplest Lagrangian that illustrates the point we wish to make:
\be
\LL = - i \hbar s \dot{\varphi} z + z^2 + f (\varphi) \; ,
\ee
where $z = \cos \theta$ and
$f (\varphi) $ is some function of $\varphi$. A typical
$z_j$-integral appearing at the $j$-th time slice in the discretized-time
version of the path integral $\int \! \! d \Omega e^{- \SS / \hbar}$ has
the form
\FL
\be
\label{eq90.1}
I = \int_{- \infty}^{\infty}  \! d \varphi_j \int_{- r}^{r} d z_j
\exp - [ - i \Delta \varphi_j z_j s + \varepsilon' z_j^2 + \varepsilon' f (
\varphi_j ) ] \; ,
\ee
\be
\label{eq90.2}
{\rm where} \; \; \dot{\varphi_j} = ( \varphi_{j} - \varphi_{j + 1} ) /
\varepsilon = \Delta \varphi_j
/ \varepsilon \; ; \qquad \varepsilon' = \varepsilon
/ \hbar \; , \qquad \varepsilon = T / (N+1) \; .
\ee
To shed light on the significance of $r$ in eq.~(\ref{eq90.1}), let us
replace the sharp cut-off at $\pm r$ by a Gaussian cut-off of width
$2r$ (the resulting change in normalization is not
relevant to our argument). Thus we replace $I$ by
\be
\label{eq91.1}
I' = \int_{- \infty}^{\infty}  \! d \varphi_j \int_{- \infty}^{\infty} d z_j
\exp - \left[ (\varepsilon' + \frac{1}{2 r^2} )
z_j^2 - i \Delta \varphi_j z_j s + \varepsilon' f (
\varphi_j ) \right] \; .
\ee
Performing the Gaussian $z_j$-integral, one readily obtains
\be
\label{eq91.3}
I' = \left( \frac{ 4 \pi}{ 2 \varepsilon' + r^{-2}} \right)^{\frac{1}{2}}
 \int_{- \infty}^{\infty} \! d \varphi_j \exp - \left[ \frac{s^2 \Delta
\varphi_j^2}{
2 (2 \varepsilon' + r^{-2} ) } + \varepsilon' f (\varphi_j) \right] \; .
\ee
Integrating out $z_j$ has thus generated a Gaussian cut-off for $\varphi_j$:
\be
\label{Brown}
\varphi_j - \varphi_{j+1} \propto  \sqrt{
2 \varepsilon' +  r^{-2}} / s \; .
\ee
If $r=\infty$, the $\varphi$-paths are Brownian-motion paths, since then
$ \varphi_j - \varphi_{j+1} \propto \sqrt{ \varepsilon}$.  However, if
$r=1$, then in the limit $\epsilon \rightarrow 0$, one has $\varphi_j -
\varphi_j \sim 1$. The coordinates on neighboring time slices therefore
can differ by an amount that is larger than infinitesimal, so that these
paths are exceedingly ill-behaved.

Instead of using a
Gaussian cut-off for the $z_j$ integral in eq.(\ref{eq91.1}),
one can use a sharp window function $w(r) = \theta(r - |
z_j |)$ as cut-off. Then the new expression (corresponding to
eq.(\ref{eq91.1})) would be exactly equal to the old eq.(\ref{eq90.1}).
The $z_j$-integral can then be performed by first
writing $w(r)$ as the Fourier integral of $\frac{\sin r q_z}{\pi q_z}$.
The calculation is more tedious, but the results are qualitatively the same.

\appendix{}

\label{appB}
This Appendix concerns  two points: \newline
(i) The identification of the prefactors of the exponentials
in the the amplitude
for a single tunneling event between neighboring minima of a periodic
potential. The relevant path  integral~(\ref{eq89.7}) need not be computed ab
initio, since one can use standard results for a particle tunneling in a
double-well potential, as presented by Coleman \cite{Col85}. \newline
(ii) Performing a summation over time histories with multiple
instantons,
to verify that  the tunnel splitting is precisely proportional
to the single-instanton amplitude.

\subsection{Single-instanton prefactor}

Our starting point is eq.~(\ref{eq91.6}) of section~\ref{phipath}:
\be
\label{eq92.2}
\LL_{\varphi} \, = \, \toh m \dot{\varphi}^2 + V(\varphi) \; ,
\ee
where
\be
\label{eq94.6}
  m \equiv (\hbar n s)^2/2a \; , \qquad   V(\varphi)=b f(\varphi) \; .
\ee
In the parlance of Rajaraman \cite{Raj82}, $\varphi$ is the coordinate of
``a particle on a unit circle'', i.e.\ all points $\varphi + 2 \pi m$ ($m$
is any integer) are physically indistinguishable.
 The potential $V (\varphi)$ is periodic and symmetric
about both $\varphi = 0$ and $\varphi = \pi / 2$ (compare
eq.~(\ref{copbarrier})):
\be
\label{copbarrier2}
V ( m \pi ) = 0 \; , \qquad
 V (\varphi) = V (- \varphi) = V (\varphi + m \pi) \quad {\mbox {\rm
for any integer $m$}} \; ,
\ee
with minima at $\varphi = \pi m$ (see fig.~\ref{barrier}).
A ``particle'' moving in this potential will execute zero-point motion, with
frequency
\be
\label{omega}
\omega = \sqrt{k/m} \; , \qquad {\rm where}\quad \left. \frac{\partial^2
V}{\partial \varphi^2} \right|_{ \varphi = m \pi} = k \; ,
\ee
around a minimum of $V(\varphi)$ for most of the time, and will occasionally
tunnel via a single-instanton event to a  neighboring minimum.
Since $\varphi$ lives on a circle, $V(\varphi)$ has only two physically
distinguishable minima, namely $| \unz \rangle \equiv \{ | 2 m \pi \rangle
\} $ and $ | \unpi \rangle \equiv \{ | (2 m +1 ) \pi \rangle \} $. They can
be associated with the two minima of a symmetric double well, since
single-instanton tunneling between neighboring minima of a periodic
potential is analogous to single-instanton tunneling in a symmetric
double-well potential.

Consider first the single-instanton tunneling events from $|\unz \rangle $
to $ | \unpi \rangle $.If the initial state is, say, $ |2 m \pi \rangle \in
|\unz \rangle $, then $ | \unpi \rangle $ can clearly be reached be either a
$(+)$-instanton or a $(-)$-instanton. The corresponding tunneling paths
$\lphi^{\pm} ( \tau)$ lead from  $ |2 m \pi \rangle \in |\unz \rangle $
 at $\tau = - T/2$ to $ |(2 m \pm 1) \pi \rangle \in |\unpi \rangle $ at
$\tau =  T/2$. Asymptotically, they have the form
\be
\label{eq92.8}
\lphi^{\pm} (\tau \rightarrow T/2)
\simeq (2 m \pm 1) \pi \mp \varphi_{\rm as}
 e^{- \omega \tau } \; .
\ee
The value of the constant
$\varphi_{\rm as}$ depends on the entire shape or $V(\varphi)$,
but is usually not too hard to obtain (see section \ref{phipath}).

Coleman shows that to lowest order in the steepest descent approximation,
the tunneling amplitude for a single $(\pm)$-instanton event has the
following form \cite{Col85appendix}:
\bea
\label{eq93.5}
\langle (2 m \pm 1 ) \pi | e^{-\hh T / \hbar} | 2 m \pi \rangle &=&
 \left( \frac{m \omega}{\pi \hbar} \right)^{\frac{1}{2}} e^{- T \omega /2} \;
T K e^{- \SSo^{\pm} / \hbar}  \; , \\
\label{eq2.9} {\rm where} \qquad K \,& = & \,
 \frac{\varphi_{\rm as}}{(\pi \hbar)^{\frac{1}{2}} } \left( \frac{ k^3}{m}
\right)^{ \frac{1}{4} } \; .
\eea
(Without loss of generality, the phases of the initial and final states have
been chosen such that $K$ is real.) For the Lagrangian~(\ref{eq92.2}),
$\SSo^+ = \SSo^-$; however, in sections~\ref{mod} and~\ref{tunnelsplit}
of this paper, $\SSo^+$ and $\SSo^-$ have different imaginary parts, hence
we keep the distinction between them in eq.~(\ref{eq93.5}).
The factor $K$ has
dimensions of inverse time and can be interpreted as an attempt frequency.
Note that it is not necessary to make a
distinction between $K_+$ and $K_-$, because to lowest order in the steepest
descent approximation, $K_+ = K_-$ (compare section~\ref{relphase}).

The total multi-instanton
{\em single}\/-instanton tunneling amplitude from $| \unz \rangle$
to $ | \unpi \rangle$ or from $ | \unpi \rangle$ to $| \unz \rangle$ is
simply the sum of the two amplitudes in eq.~(\ref{eq93.5}):
\bea
 \nonumber
\langle \unz | e^{-\hh T / \hbar} | \unpi \rangle &=&
\langle \unpi | e^{-\hh T / \hbar} | \unz \rangle \\
\label{eq93.5a}  &=&
\left( \frac{m \omega}{\pi \hbar} \right)^{\frac{1}{2}} e^{- T \omega /2}
T K( e^{- \SSo^{+} / \hbar} + e^{- \SSo^{-} / \hbar} ) \; .
\eea

\subsection{Sum of multi-instanton paths}

The total amplitude for tunneling from $| \unz \rangle $ to
$| \unpi \rangle$ can be written as an infinite series, the $n$-th term of
which represents a string of $n$
single-instanton events, widely separated in time from
each other (the ``dilute gas approximation").
Letting the initial and final states be
separated by a time interval $T$, with $T \rightarrow \infty$,
the net amplitude is \cite{Col85}
\bea
\nonumber
\langle \unpi | e^{ - \! \whh T / \hbar} | \unz \rangle
&=&  \left( \frac{m \omega}{\pi \hbar}
\right)^{\frac{1}{2}}
e^{- T \omega /2}
\sum_{n= {\rm odd}}
{\textstyle \frac{1}{n !}} [ T K( e^{- \SSo^{+} / \hbar} + e^{- \SSo^{-} /
\hbar} )]^{n} \; \\
\label{eq93.5b}
&=& \left( \frac{m \omega}{\pi \hbar} \right)^{\frac{1}{2}} e^{- T \omega /2}
\bigl( e^{ T \Delta / ( 2 \hbar)} - e^{- T \Delta  / ( 2 \hbar)} \bigr)\; ,
\eea
\be
\label{eq93.8}
{\rm where} \quad
\Delta / 2  \, = \, \hbar K \left( e^{- \SSo^+ /  \hbar}
+ e^{- \SSo^{-} / \hbar} \right) \; .
\ee

In the limit $T \rightarrow \infty$, the lowest energy eigenvalues
can be read off from the general expression
\be
\label{eq94.2}
\langle  \unpi | e^{ - \! \whh T / \hbar} | \unz \rangle \, = \,
\sum_{n} e^{- E_n T / \hbar} \langle \unpi | n \rangle \langle n | \unz \rangle
\; .
\ee
This gives
\be
\label{eq94.2a}
E_{\pm} = \hbar \omega / 2 \pm \Delta / 2
\ee
for the energies of
the lowest two energy states of the double-well problem.
The first term is obviously the usual zero-point energy.

Substituting eqs.~(\ref{eq2.9}) and
eq.~(\ref{eq94.6}) into eq.~(\ref{eq93.8})
gives eq.~(\ref{eq95.7}), as was to be shown.

\appendix{}

\label{appC}
{\em Theorem:}

Consider any coplanar
classical ground state configuration of the Kagom\'e antiferromagnet. Choose
any closed loop of alternating AB-spins.
The total number of spins on such a loop
is always of the form $4n +2$, where $n$ is an integer.

{\em Proof:}
The Bravais lattice of the Kagom\'e net, which has one site
 located at the center of every Kagom\'e hexagon  (marked
X in fig.~\ref{app1}),
forms a triangular ``superlattice'' (SL). Our proof is formulated in terms of
the SL, and is divided into four steps.

\subsection{Tiling by rhombi}

In any coplanar classical ground state, every
Kagom\'e-triangle must contain exactly one spin of each of the types A,
B and D. This  can be used to construct a tiling of the Kagom\'e lattice by
decorated rhombi, according to the following scheme:

 At the center of every ``bond" connecting two SL nearest neighbor sites
(i.e. Kagom\'e  vacancies) there is a Kagom\'e spin, and every
Kagom\'e spin is at the
center of a SL bond. Any rhombus made up of two neighboring
SL triangles thus is decorated by a bowtie of five spins, one at its center
and four at the centers of its four sides. These five spins form two
head-to-head Kagom\'e-triangles. It is clear that for any given classical
ground state, the SL can be tiled
by decorated rhombi according to the following matching rules:
\newline
(a) On every rhombus, the two bowtie-triangles must each
contain exactly one spin of each of the types A, B and C
(see fig.\ref{app1}).
\newline
(b) The shared side of any two neighboring rhombi must have
 matching spins.

\subsection{Definition of cluster}

Associate a connected ``cluster" of SL-bonds
with the given AB-loop by connecting all SL-sites on the inside of the
AB-loop to their nearest neighbors (fig.~\ref{cluster}).
The smallest cluster possible is just a single SL-site; the associated
AB-loop surrounding it is the hexagon of six spins studied in this paper.
The next-smallest cluster consists of two SL-sites connected by a
SL-bond; the associated AB-loop has ten spins.

A cluster that can not be separated into two disconnected clusters
by cutting a single SL-bond will be called a ``blob".
The smallest blob possible is just a single SL-site. The
next-smallest blob is a SL-hexagon (a SL-triangle is not allowed by
the tiling rules).
A large cluster may contain many blobs, say $m$, connected
 to each other by $m-1$ single bonds, to be called ``cutting links"
(fig.~\ref{cluster}). At the center of each cutting link there will necessarily
always be a $C$-spin (because both sides of the cutting link are lined
by AB's).

\subsection{Reducing clusters to blobs}
To count the number, say $M$,
of AB's of the original loop that surrounds (and
defines) the cluster, we systematically cut cutting links to break down the
cluster into disconnected blobs:
Cut a cutting link, thus separating the mother cluster into
two disconnected daughter clusters. Close the AB-loop around each
daughter cluster by replacing the $C$-spin of the freshly
cut cutting link
by an A-spin for one daughter and a B-spin for the other, whichever
is appropriate to maintain the alternation of A's and B's around each
daughter (the reason that one can replace one $C$-spin by two
different spins (A and B) is that the two daughter clusters are considered
to be separated from each other once the link has been cut).
Then we have $M = D_1 + D_2 - 2$, where $M$, $D_1$ and $D_2$
are the number of AB-spins surrounding the mother cluster and the
two daughter clusters respectively.
By repeating this process until all cutting links have been cut,
we are left with $m$ ``disconnected blobs", each surrounded by an
AB-loop containing, say,  $B_i$ spins, where $i = 1, \dots, m$. Thus,
\be
\label{daughter}
M = \sum_{i=1}^m B_i - 2 (m-1)
\ee

\subsection{Perimeter of blobs}

Next we show that $B_i$ always has the form $B_i = 4n + 2$.
The smallest disconnected blob, namely the single SL-site, has $B= 6$.
Any larger disconnected
blob can be tiled by decorated rhombi, in such a way that there are only
C-spins at the centers of the bonds that form the perimeter of the blob,
because the disconnected blob is surrounded by a loop of AB-spins.
 Any closed shape tiled by rhombi always has an
even number, say $2S_i$, of perimeter bonds, and hence of perimeter
SL-sites. Any perimeter SL-site can be associated with either 1, 2, or 3
AB-spins on its outside, depending on whether the perimeter turns
outward, keeps straight or turns inward at that site (fig.~\ref{cluster}).
There are exactly six more inward-turning than outward-turning
turns around the perimeter of any blob
 (because at each turn the direction of the perimeter changes
by $60^{\circ}$, since $120^{\circ}$ turns are not allowed by the tiling
rules). Therefore, the net number of AB-spins around a blob always has the
form $B_i = 2 (2S_i) + 6$. Inserting this result into eq.(\ref{daughter}),
we find
\be
\label{mother}
M = 4 \left( \sum_{i = 1}^m  S_i + 1 \right) \, + \, 2 \; .
\ee
This proves the theorem.

\figure{The \sqiii\ ground state of the Kagom\'e lattice.
The letters A, B and C represent the spin directions illustrated in
fig.~\ref{n,triangle}b. The symbols P and Q and the labels 0 to 5
are the ones used in appendix~\ref{appD}. \label{ABABAB}}

\figure{(a) The unit vector ${\mbox{\boldmath $n $}} = (\phi, \theta)$. (b)
 Type A, B and C  spins on a triangle of the Kagom\'e lattice. The common
 plane of
A, B and C, i.e.\ the ``reference plane'', defines the $\varphi = 0$ and
$\varphi = \pi$ directions in (a).  \label{n,triangle}}

\figure{The ``weathervane mode''. All six spins on an ABABAB-hexagon
rotate collectively, as a rigid unit in spin space, by $180^{\circ}$
around the C-direction, and end up forming a BABABA-hexagon.
The labels 0 to 5 correspond to the index $l$ in eqs.~(\ref{AFM})
and (\ref{cop}).  \label{BABABA}}

\figure{A typical shape function $f(\varphi)$ for the coplanarity potential
$\hh_{\rm cop}$. The function shown here corresponds to the $s=1$ case
of the function calculated
in appendix~\ref{appD}, eq.~(\ref{hcopf}), and used in sections~\ref{phipath}
and~\ref{Kag}. The arrow shows the position of $\past$, which marks the
crossover of $f(\varphi)$ from quadratic to linear behavior.
However, the general discussion of sections~\ref{mod}
and~\ref{tunnelsplit}, up to eq.~(\ref{eq92.8a}), is independent of the
particular shape of  $f(\varphi)$, and only requires the symmetries
of eq.~(\ref{copbarrier}) to hold.
 \label{barrier}}

\figure{Contours of constant $\hh (\varphi, z)$. In (a), Im$[z] = 0$, and
the contours in the $(\varphi, {\rm Re}[z])$-plane
depict ``valleys'' centered
 at $(\varphi,z) = (0({\rm mod} \pi), z_g)$. In (b), Re$[z] = z_g$, and
the contours in the $(\varphi, {\rm Im}[z])$-plane show ``ridges''
centered at $(\varphi, z) = (\pm  \pi/2 ({\rm mod} \pi ), z_g )$.
 The $(+)$-instanton and $(-)$-instanton tunneling paths are
contours of $\hh = 0$ around these ridges.
\label{contours}}

\figure{An $(+)$-instanton $\lphi^+(\tau)$ followed by two $(-)$-instantons
$\lphi^-(\tau)$, where $\tau$ is the (imaginary) time parameter.
\label{instantons}}

\figure{The centers of Kagom\'e hexagons (marked X) form a triangular
 superlattice (SL). Connecting these by the
dashed lines divides the plane into (large) rhombi,
each with, say, an A-spin at its center, surrounded by four other sites in a
``bow-tie'' shape.   \label{app1}}

\figure{An AB-loop surrounding a cluster of bonds (heavy lines) on the
superlattice.
A- and B- spins on this loop are represented by heavy dots and squares
respectively.
There is one large blob at the center, whose interior is divided into
the twelve large triangles.
Each of the four outside vertices
is a blob consisting of just one SL-site
(they are the vertices with just one or three neighbors),
and each of the four outside bonds
(which connect these blobs to each other and to the large blob)
is a cutting link.
The (perimeter) of the large blob has eight inward and two outward turns.
  \label{cluster}}


\begin{thebibliography}{WWW99}
\bibitem{nneigbor} The second- and third- nearest neighbors of a site
are separated  from the site by $\sqrt{3} a_o$ and $2 a_o$ respectively.
\bibitem{moregeneral} The manipulations we employ in section~\ref{Kag} also
work, however, if one chooses a more general functional form for $\hh_{\rm
cop}$, e.g. $\frac{1}{6} J_b \sum_{l=0}^5 f( \frac{1}{2}( \varphi_l +
\varphi_{l+1} ) + \pi)$.
\bibitem{multinst} This result persists when multiple
$(+)$- and $(-)$-instanton
events are considered (see appendix~\ref{appB}).
\bibitem{boundaries} Since the Lagrangian is of first order in
time derivatives, care has to
be exercised when specifying the boundary conditions for a classical path,
since specifying
the initial and final values of both $\varphi$ and $z$ in general
leaves one with an overdetermined set of equations \cite{Kla79,KG92}. In
the problem considered in this paper, however, this does not cause
any complications, since the boundary conditions for the tunneling
paths of interest to us happen to be such that solutions to the
equations of motion do exist.
\bibitem{diagonalize}  Note that $\delta^2 {\cal L}$ is still
not totally diagonal
in the index $l$, since $\whdz_l$ is coupled to $\whdz_{-l}$,
and $\whdphi_l$ to $\whdphi_{-l}$.
If desired, a further change of variables to a set of purely real
coordinates may be performed to completely diagonalize $\delta^2 \LL$ with
respect to the index $l$. However, for our purposes, this is not
necessary.
\bibitem{varwave} This is equivalent to using a variational wavefunction
and assuming that in terms of the new coordinates, it has the same six-fold
symmetry that the system had in terms of the old coordinates when $\varphi =
0$.
\end{thebibliography}
\end{document}